\documentclass[11pt,a4paper]{article}
\pdfoutput=1
\usepackage{jheppub}
\usepackage{amsfonts,amssymb,amsmath}
\usepackage{epsfig, graphicx,yfonts}

\newcommand{\be}{\begin{equation}}
\newcommand{\ee}{\end{equation}}
\newcommand{\bea}{\begin{eqnarray}}
\newcommand{\eea}{\end{eqnarray}}

\newcommand{\mt}[1]{\textrm{\tiny #1}}

\newcommand{\pa}{\partial}

\newcommand{\vev}[1]{\langle #1\rangle}

\def\nc {N_\mt{c}}

\def\uh {u_\mt{H}}

\def\lgb {\lambda_\mt{GB}}
\newcommand{\cs}{{\cal S}}

\newcommand{\vard}[2]{\frac{\delta #1}{\delta #2}}

\title{Holographic renormalization and \\ anisotropic black branes in higher curvature gravity}
\author{Viktor Jahnke, Anderson Seigo Misobuchi, and Diego Trancanelli} 
\affiliation{Institute of Physics, University of S\~ao Paulo\\ 05314-970 S\~ao Paulo, Brazil} 

\abstract{
We consider five-dimensional AdS-axion-dilaton gravity with a Gauss-Bonnet term and find a solution of the equations of motion which corresponds to a black brane exhibiting a spatial anisotropy, with the source of the anisotropy being an axion field linear in one of the horizon coordinates. Our solution is static, regular everywhere on and outside the horizon, and asymptotically AdS. It is analytic and valid in a small anisotropy expansion, but fully non-perturbative in the Gauss-Bonnet coupling. We discuss various features of this solution and use it as a gravity dual to a strongly coupled anisotropic plasma with two independent central charges, $a\neq c$. In the limit of small Gauss-Bonnet coupling, we carry out holographic renormalization of the system using (a recursive variant of) the Hamilton-Jacobi method and derive a generic expression for the boundary stress tensor, which we later specialize to our solution. Finally, we compute the shear viscosity to entropy ratios and conductivities of this 
anisotropic plasma.   
}
  
\keywords{Gauge-gravity correspondence, Holography and quark-gluon plasmas}  

\emailAdd{viktor.jahnke} 
\emailAdd{anderson.misobuchi} 
\emailAdd{dtrancan@usp.br} 
  
\begin{document}  

\maketitle
\setlength{\parskip}{3pt}


\section{Introduction}

The AdS/CFT correspondence \cite{Maldacena:1997re,duality2,duality3} represents a remarkable tool in the study of the strongly coupled gauge theories which can be mapped to a dual, weakly coupled gravitational description. The correspondence is best understood, and mostly used, in the limit in which both $N$ and $\lambda$, the rank of the gauge group and the 't Hooft coupling of the gauge theory, respectively, are infinite. This maps to the classical supergravity limit of the string side. Investigating departures from this limit implies introducing $\alpha'$ and loop corrections for the string and it is clearly of the utmost importance for a series of reasons, from achieving a deeper understanding of how the correspondence works in larger regions of the parameter space, to modeling more realistic gauge theory systems, where $N$ and $\lambda$ are obviously not infinite. A systematic treatment of $\alpha'$ and loop corrections is however a tall order, given their complexity. For example, the leading finite 
coupling corrections to type IIB supergravity arise as stringy corrections with schematic form $\alpha'^3 R^4$. 

One more modest approach is to consider simple generalizations of Einstein gravity, where higher curvature corrections are under control and calculable, in the hope to gain some qualitative understanding of the effects they might have and, perhaps, uncover some universal properties. A well-studied family of corrections is represented by Lovelock theories of gravity \cite{lovelock1,lovelock2,lovelock3,lovelock4}.\footnote{Reviews on Lovelock theories with an emphasis on their relevance in the AdS/CFT context can be found in, e.g., \cite{reviewlovelock0,reviewlovelock,reviewlovelock2}.} These theories are defined as natural extensions of the Einstein-Hilbert action to dimensions higher than four. The main characteristic of Lovelock gravities is the fact that, albeit being defined in terms of higher curvature/derivative terms, they yield second order equations of motion and are free of pathologies. The first Lovelock correction, which is present already in five dimensions, is given by the Gauss-Bonnet (GB) 
term, which is quadratic in the curvature. Another point of interest in this correction is that, besides being calculable, it possesses a wealth of exact black hole solutions with AdS asymptotics; see e.g. \cite{branches,camanhoetal} and \cite{bestiary} for a comprehensive review.

It is clearly worthwhile to try to find as many new solutions as possible and increase the arena of models where explicit computations can be performed. With this motivation in mind, in this paper we consider a GB correction to Einstein-Hilbert gravity in five dimensions with a negative cosmological constant and a coupling to an axion-dilaton field. It is not clear whether this system might be obtained by some string theory compactification, so that our philosophy in this work is `bottom-up'.

The first result of our analysis is a new solution of the equations of motion representing a black brane with a translationally invariant but anisotropic horizon. The force responsible for keeping the horizon in an anisotropic state is furnished by the axion field, which we take to have a fixed profile in the radial coordinate but to depend linearly in one of the horizon coordinates. This is similar to what has been done in \cite{ALT} and later in \cite{MT,MT2}. This new solution is interesting from a purely General Relativity point of view, for it opens up the possibility to study the thermodynamics of a black brane which depends on several parameters (the temperature, the GB coupling and an anisotropy parameter), presumably giving rise to a rich phase space. In this paper we move a first step toward the study of such thermodynamics by computing the boundary stress tensor, which represent our second and perhaps most important result. This computation requires the machinery of holographic renormalization. More specifically, we 
use a Hamiltonian approach to the problem, rather than the more commonly used Lagrangian one, in the incarnation of the recursive Hamilton-Jacobi method developed in \cite{Yiannis} for the AdS-Einstein system with axion-dilaton (without higher derivative corrections). Holographic renormalization of Einstein gravity with the GB term, but without any other field turned on, has been performed in \cite{LiuSabra}.

A more applied motivation for our work is given by the study of the quark-gluon plasma (QGP) produced in the ultra-relativistic collision of heavy ions at RHIC \cite{rhic,rhic2} and LHC \cite{lhc}. Contrary to na\"ive expectations, this plasma turns out to be a strongly coupled fluid \cite{fluid,fluid2}, rather than a quasi-ideal free gas of quarks and gluons. This fact renders a perturbative approach of limited applicability and motivates the use of the AdS/CFT correspondence; see \cite{reviewmateosetal} for a review of applications of AdS/CFT to the study of the QGP. One of the diagnostics of the strongly coupled nature of this fluid is represented by `elliptic flow', i.e. the anisotropic evolution of the fluid in the initial stages before isotropization. Recently, there has been some interest in modeling this anisotropy at strong coupling \cite{MT,MT2} and in studying how various observables may be affected by it. Some of the studies that have been performed include the computation of the shear viscosity to entropy density ratio \cite{Rebhan,mamo}, the drag force experienced by a heavy quark \cite{Chernicoff:2012iq,giataganas,new_drag}, the energy lost by a quark rotating in the transverse plane \cite{fadafan}, the stopping distance of a light probe \cite{stopping}, the jet quenching parameter of the medium \cite{giataganas,Rebhan:2012bw,jet}, the potential between a quark and antiquark pair, both static \cite{giataganas,Rebhan:2012bw,Chernicoff:2012bu,indians} and in a plasma wind \cite{Chernicoff:2012bu}, including its imaginary part \cite{Fadafan:2013bva}, Langevin diffusion and Brownian motion \cite{langevin,langevin2,langevin3}, chiral symmetry breaking \cite{Ali-Akbari:2013txa}, the production of thermal photons \cite{photon,Wu:2013qja,Arciniega:2013dqa} and dileptons \cite{dileptons}, and the introduction of a chemical potential \cite{chempot,chempot1}; see \cite{Giataganas:2013lga} for a review of some of these computations and \cite{seealso} for similar computations in a fluid with dilaton-driven anisotropy. 

In order to achieve a more realistic model of the anisotropic plasma it is obviously important to relax some of the assumptions (like the infinite coupling and infinite number of colors) that go into the simplification of having a classical gravity dual. The GB coupling that we introduce here corresponds to allowing for different central charges, $a\neq c$, in the gauge theory \cite{aneqc,aneqc1,aneqc2}. We compute these two central charges for our particular solution, verifying that they are indeed different. On general grounds, looking at how higher derivative terms affect physical observables on the gauge theory might also be useful to constrain the string landscape, e.g. by excluding regions of parameters that would result in pathologies, as advocated for example in \cite{camanhoetal,maldaedel}. As a final, concrete application of our geometry we compute the shear viscosity over entropy density ratio (in a few, equivalent ways) and the plasma conductivities.

This paper is organized as follows. In Sec.~\ref{sec2} we present our solution and compute its temperature and entropy density. In Sec.~\ref{sec3} we carry out the holographic renormalization, obtaining general formulas for the expectation values of the stress tensor and of the axion and dilaton operators. In Sec.~\ref{sec4} we specialize those formulas to the case at hand and discuss the various features of energy density and pressures. In Sec.~\ref{sec5} we use the solution as a model for a strongly coupled anisotropic plasma and compute various transport coefficients of interest, such as the shear viscosity over entropy ratios, both along the anisotropic direction and in the transverse plane, using the membrane paradigm. We finally discuss our results and outline possible future extensions of our work in Sec.~\ref{sec6}. A series of appendices contains some of the more technical details of our computations, like the explicit derivation of the solution and the derivation of the shear viscosity tensor using 
alternative methods.


\section{Action and solution}
\label{sec2}

We are interested in five-dimensional gravity with a negative cosmological constant and the inclusion of a Gauss-Bonnet term, which we also couple to an axion-dilaton system in the following way
\begin{equation}
S=\frac{1}{16 \pi G}\int d^5x\, \sqrt{-g}\left[R+\frac{12}{\ell^2}-\frac{1}{2}(\partial \phi)^2-\frac{e^{2\phi}}{2}(\partial \chi)^2+\frac{\ell^2}{2}\lgb{\cal L}_\mt{GB}\right]+S_\mt{GH}\,.
\label{action1}
\end{equation} 
The scalar fields $\phi$ and $\chi$ are the dilaton and axion, respectively, $\lgb$ is the (dimensionless) Gauss-Bonnet coupling and 
\begin{equation}
{\cal L}_\mt{GB}=R^2-4 R_{mn}R^{mn}+R_{mnrs}R^{mnrs}
\label{GBterm}
\end{equation}
is the Gauss-Bonnet term. $\ell$ is a parameter with dimensions of length that we set to one in what follows, without loss of generality. We use the Latin indices $m,n,\ldots$ for the five-dimensional coordinates $(t,x,y,z,u)$, with $u$ being the radial coordinate. The term $S_\mt{GH}$ is the usual Gibbons-Hawking term, necessary to render the variational problem well posed. When $\lgb=0$ the action above can be obtained from type IIB superstrings \cite{MT,MT2}, but this is no longer true when the Gauss-Bonnet coupling is turned on. In fact, it is not clear whether (\ref{action1}) can be obtained from any string theory compactification, so that our point of view in the present paper is `bottom-up', as already discussed in the Introduction.

The field equations for the metric resulting from the action above are given by
\be
R_{mn}-\frac{1}{2}g_{mn}R+\frac{\lgb}{2} \delta {{\cal L}_\mt{GB}}_{mn}=
\frac{1}{2}\partial_m \phi\, \partial_n \phi+\frac{1}{2}e^{2\phi} \partial_m \chi \,\partial_n \chi
-\frac{g_{mn}}{4}\left[(\partial \phi)^2 +e^{2\phi}(\partial \chi)^2-12 \right]
\,,
\label{eom}
\ee
where
\be
 \delta {{\cal L}_\mt{GB}}_{mn} =-\frac{g_{mn}}{2}{\cal L}_\mt{GB}-4R_m^{\,\,\,\,\,r} R_{rn}+2 R_{mn}R - 4 R^{rs}R_{mrns}+2R_m^{\,\,\,\,\,rst}R_{nrst}
\ee
is the variation of the Gauss-Bonnet term. The equations for the dilaton and axion read instead
\bea
 \partial_m(\sqrt{-g}g^{mn}\partial_n\phi)=\sqrt{-g}e^{2\phi}(\partial\chi)^2\,, 
\qquad  \partial_m(\sqrt{-g}e^{2\phi}g^{mn}\partial_n\chi)=0\,.
\label{eom-dil}
\eea
We want to obtain a solution which displays a spatial anisotropy. This is achieved by singling out one direction, say the $z$-direction, which will be later identified with the `beam direction' in a heavy ion collision experiment occurring in the boundary theory. To get an anisotropy between the $z$-direction and the $xy$-directions (the transverse plane to the beam), we consider the following Ansatz\footnote{Note that this Ansatz is slightly different than the one used in \cite{MT,MT2}.}
\be
ds^2=\frac{1}{u^2}\left( -F B\, dt^2+dx^2+dy^2+H\, dz^2 +\frac{du^2}{F}\right).
\label{ansatz-metric}
\ee
All the metric components $F$, $B$, and $H$, as well as the dilaton $\phi$, depend solely on the radial coordinate $u$. This guarantees that the solution be static. In this parametrization the boundary is located at $u=0$. $F$ is a `blackening factor' that introduces an horizon in the geometry at $u=\uh$, where $F(\uh)=0$. There is a scaling symmetry in the coordinates $t$ and $z$ that allows us to set $B_\mt{bdry}F_\mt{bdry}=H_\mt{bdry}=1$, thus recovering a canonically normalized AdS metric in the UV region (with radius $1/\sqrt{F_\mt{bdry}}$). Here and in what follows we use the subscript `bdry' to denote the value of the fields at $u=0$. 

Following \cite{ALT,MT} we consider an axion field which has a constant profile in the radial direction and depends linearly on $z$
\be
\chi=a \, z\,.
\ee
The parameter $a$ has dimensions of energy and controls the amount of anisotropy. It is clear that this is a solution of the axion equation, since the metric is diagonal and the metric and dilaton do not depend on $z$.

In this paper we limit ourselves to considering the case of small anisotropy, which will allow for an analytic solution of the equations of motion. To do this we expand all the fields around the (isotropic) Gauss-Bonnet black brane solution\footnote{See e.g. \cite{Brigante} or \cite{reviewlovelock} for a review.}
\bea
\phi(u)&=&a^2 \phi_{2}(u)+O(a^4)\,,\cr
F(u)&=&F_0(u)+a^2  F_{2}(u)+O(a^4)\,,\cr
B(u)&=&B_0\left(1+a^2 B_{2}(u)+O(a^4)\right)\,,\cr
H(u)&=&1+a^2  H_{2}(u)+O(a^4)\,,
\label{ansatz-sol}
\eea
where
\be
F_0(u)=\dfrac{1}{2\lgb}\left(1-\sqrt{1-4\lgb\left(1-\frac{u^4}{\uh^4}\right)}\right)\,,\qquad \lgb<\frac{1}{4}\,.
\label{F0}
\ee
This is a solution of the equations of motion when $a=0$. In order to have a unit speed of light at the boundary we set
\be
B_0=\frac{1}{2}\left(1+\sqrt{1-4\lgb}\right)\,.
\ee
This is possible due to the scaling symmetry in $t$, which we have mentioned above. Note that only even powers of $a$ can appear in the expansion because of the symmetry $z\to-z$. 

\begin{figure}
\begin{center}
\setlength{\unitlength}{1cm}
\includegraphics[scale=.75]{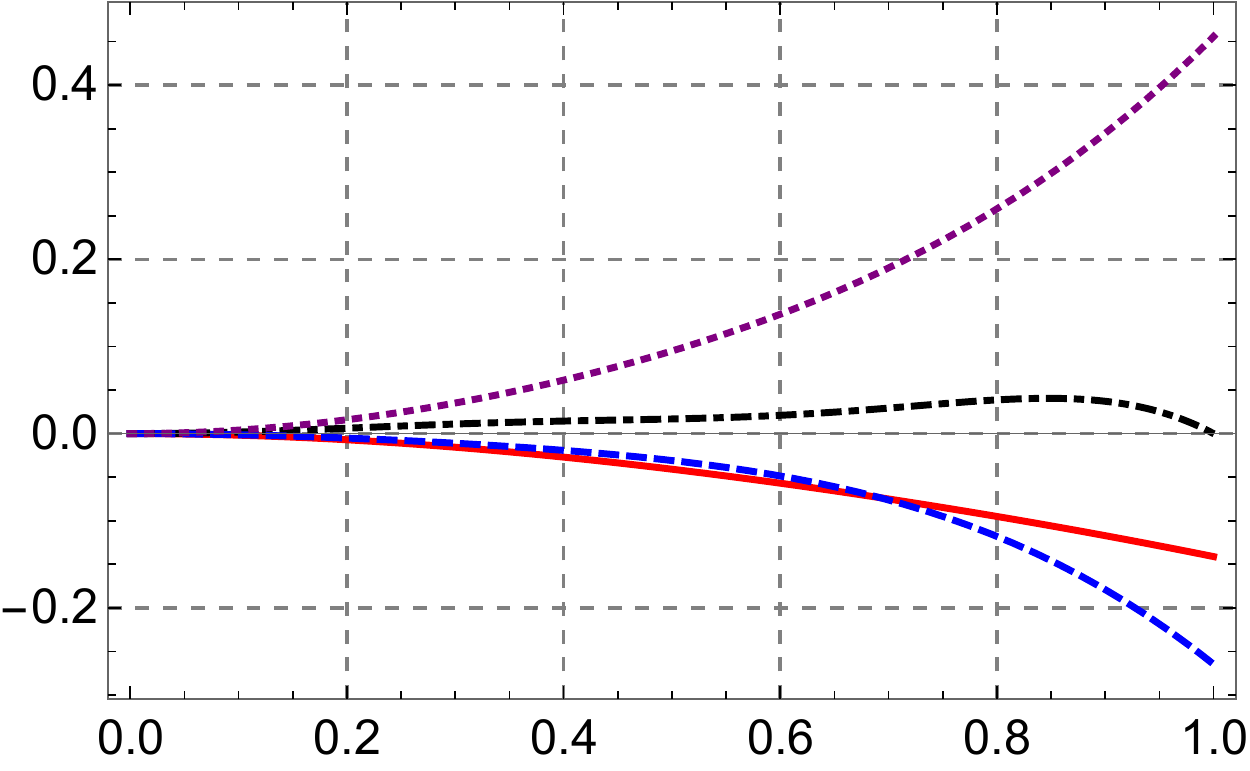} 
\put(-1.4,-.3){$u/\uh$}
\put(-2.9,4.4){$H_2$}
\put(-1.6,3.2){$F_2$}
\put(-1.4,.8){$B_2$}
\put(-1.2,2.1){$\phi_2$}
\end{center}
\caption{\small
The metric functions at order $O(a^2)$. Here we have set $\lgb=0.2$.}
\label{fig-metric}
\end{figure}
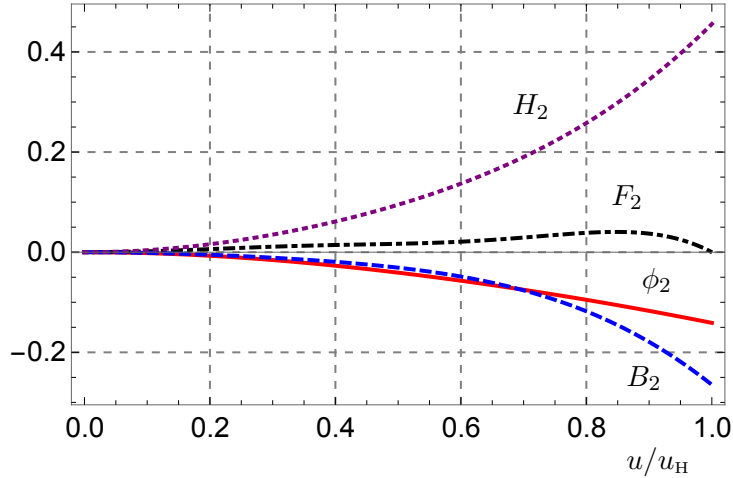

Luckily it is possible to solve the equations analytically at order $O(a^2)$. The equations at this order and the explicit solutions are detailed in App.~\ref{app1}. A plot of representative solutions is contained in Fig.~\ref{fig-metric}, where the regularity of the geometry is explicitly exhibited. 
Here we just mention that we have fixed the integration constants in such a way that all the metric functions are regular at the horizon and moreover
\be
\phi_{2,\mt{bdry}}=F_{2,\mt{bdry}}=B_{2,\mt{bdry}}=H_{2,\mt{bdry}}=0\,,
\ee
thus recovering AdS in the UV. A direct computation of the Kretschmann invariant $R_{mnpq}R^{mnpq}$ shows no singularity in the geometry except for $\lgb=1/4$, which is however excluded, as can be seen from (\ref{F0}). 

Unfortunately, we have not been able to find analytic solutions beyond order $O(a^2)$ and most likely a numerical analysis will turn out to be necessary to go to higher anisotropies. This is however beyond the scope of the present paper. It should be possible, in principle, to consider arbitrarily large values of $a$, as in the pure Einstein-Hilbert case of \cite{MT,MT2}.

The temperature of the solution can be computed as usual from the standard requirement that the (Euclideanized) metric be regular at $u=\uh$. One finds that 
\bea
T=-\frac{F'(u)\sqrt{B(u)}}{4\pi}\Big|_{u=\uh}\,.
\eea
Specializing to our solution this becomes\footnote{Note that this expression is valid, and real, even for negative $\lgb$.}
\bea
T&=&\sqrt{B_0}\left(\frac{1}{\pi\uh}-\frac{2B_0-6\lgb+\sqrt{\lgb}\log\left(\frac{1+2\sqrt{\lgb}}{1-2\sqrt{\lgb}}\right)
-\log\left(\frac{4B_0}{\sqrt{1-4\lgb}}\right)}{48\pi(1-4\lgb)}\uh a^2+O(a^4)\right)\,.
\cr&&
\label{temppos}
\eea
This equation can be easily inverted to find $\uh$ as a function of $T$. 

For planar black holes in GB gravity the entropy density is still given by the usual formula in terms of the area of the horizon. We find (here $V_3$ is the infinite volume $\int dx\, dy\, dz$)
\bea
s=\frac{A_\mt{hor}}{4 G V_3}
=\frac{\pi}{4GB_0^{3/2}}\left(\pi^2 T^3+\frac{1}{8}T B_0 a^2+O(a^4)\right)\,.
\label{entropy}
\eea
We notice that for $\lgb=0$ this matches the result obtained in \cite{MT,MT2}.

A final comment on the IR behavior of the geometry is in order. The solution of \cite{MT,MT2} was interpolating between AdS boundary conditions in the UV and a Lifshitz-like scaling solution \cite{ALT} in the IR. We believe that the finite $\lgb$ generalization discussed here does not share this feature with \cite{MT,MT2}, although we have not been able to prove this rigorously. More specifically, we have not been able to find a scaling solution in the IR (even for $T=0$), as done in \cite{ALT} for the case $\lgb=0$. One obstruction might be that Lifshitz solutions in GB gravity seem to require to tune the cosmological constant in ways that are not compatible with our equations. For example, in the case of GB gravity coupled to a massive vector field the condition for a Lifshitz scaling is that the cosmological constant be half of the usual value \cite{Mann}.\footnote{A flow between a Lifshitz solution in the UV and an AdS solution in the IR for GB-gravity coupled to a massive vector field was discussed in \cite{ParkMann}.}  It would certainly be interesting to settle this point, but this goes beyond the scope of the paper.


\section{Holographic renormalization}
\label{sec3}

In this section we use holographic renormalization techniques to compute the 1-point function of the boundary stress tensor associated to our gravitational system; see \cite{reviewHR} for a review. In fact, we consider a generalization of (\ref{action1}), where the coefficient of the axion kinetic term is allowed to be a generic function $Z(\phi)$ of the dilaton. We also maintain the metric and axion-dilaton generic. We use the recursive Hamilton-Jacobi method that was introduced in \cite{Yiannis} for axion-dilaton gravity without Gauss-Bonnet term. Our main result are the expressions (\ref{stress-final})-(\ref{chi-final}) for $\vev{T_{ij}}$, $\vev{{\cal O}_\phi}$ and $\vev{{\cal O}_\chi}$, which are general and which we specialize to our solution (\ref{ansatz-sol}) in the next section.

\subsection{Metric and action in ADM form}

The recursive Hamilton-Jacobi method of \cite{Yiannis}, which we follow closely in this section, makes use of the ADM formalism, in which a manifold $\mathcal{M}$ is foliated with hypersurfaces $\Sigma_r$ of constant radial coordinate, which we call $r$ in this section. In this coordinate, which plays the role of Hamiltonian time, the boundary is located at $r=\infty$. The metric on $\mathcal{M}$ takes the form
\begin{equation}
 ds^2=(N^2+N_iN^i)dr^2+2N_idrdx^i+\gamma_{ij}dx^idx^j\,,
\end{equation}
where $N$ and $N_i$ are the lapse and shift function, respectively, and $\gamma_{ij}$ is the induced metric on $\Sigma_r$. We use the Latin indices $i,j,\ldots$ to label the coordinates $(t,x,y,z)$ on $\Sigma_r$. In terms of these fields, the extrinsic curvature is given by
\begin{equation}
 K_{ij}=\frac{1}{2N}(\dot{\gamma}_{ij}-D_iN_j-D_jN_i)\,,
\end{equation}
with the dot denoting differentiation with respect to $r$ and $D_i$ being the covariant derivative associated to $\gamma_{ij}$.

The axion-dilaton part of (\ref{action1}), without Gauss-Bonnet contribution, turns out to be given by\footnote{We gauge-fix $N=1$ e $N_i=0$ and set, for this section, $\frac{1}{16\pi G} =1$. Note that we use a different normalization (a factor of $1/2$) in our scalar kinetic terms compared to the scalar kinetic terms in \cite{Yiannis}.\label{shiftnote}}
\bea
 S_\mt{axion-dilaton} &=& \int_{\mathcal{M}}d^dx\sqrt{-g}\Big[\mathcal{R}+K^2-K_{ij}^2+(d-1)(d-2)
 \cr &&\hskip2.5cm  
-\frac{1}{2}\left(\dot{\phi}^2+Z(\phi)\dot{\chi}^2+\gamma^{ij}(\partial_i\phi\partial_j\phi+Z(\phi)\partial_i\chi\partial_j\chi)\right)\Big]\,.
\eea
Here and in the following we leave the function in the axion kinetic term as a generic function of the dilaton, $Z(\phi)$. Later we will specialize to $Z(\phi)=e^{2\phi}$ and to $d=5$, which is the case considered in the previous section. We denote with the calligraphic fonts $\mathcal{R}$, $\mathcal{R}_{ij}$, etc. the curvature on $\cal{M}$ computed in terms of $N,N_i$ and $\gamma_{ij}$. All the contractions of the $i,j,\ldots$ indices are performed with $\gamma^{ij}$. The Gauss-Bonnet contribution is (see for example eq. (2.8) of \cite{LiuSabra})
\bea
 S_\mt{GB} &=&\frac{1}{2}\int_{\mathcal{M}}d^dx\sqrt{-g}\left[(\mathcal{R}^2+K^2-K_{ij}^2)^2-4(\mathcal{R}_{ij}+KK_{ij}-K_{ik}K_j^k)^2
 \right.\nonumber\\
 && \hskip 3cm 
 +(\mathcal{R}_{ijkl}+K_{ik}K_{jl}-K_{il}K_{jk})^2
 -\frac{4}{3}K^4+8K^2K_{ij}^2
 \nonumber\\&&\hskip 4cm 
 \left. -\frac{32}{3} KK_i^jK_j^kK_k^i 
 -4(K_{ij}^2)^2+8K_i^jK_j^kK_k^lK_l^i\right]\,,
\eea
with $K=\gamma^{ij}K_{ij}$. The Gibbons-Hawking terms have already been included in the actions above, but they get canceled by boundary terms coming from the bulk actions. The total action is then
\begin{equation}
 S=S_\mt{axion-dilaton}+\lgb S_\mt{GB}\,.
 \label{total-action}
\end{equation}


\subsection{Radial evolution Hamiltonian}

The next ingredient in the algorithm is to compute the Hamiltonian for radial evolution, which is associated to the Lagrangian $L$ defined by $S=\int dr L$. To this scope, we need the canonical conjugate momenta
\bea
\label{eq:canonicalmomentum}
 \pi^{ij}&\equiv &\frac{1}{\sqrt{-\gamma}}\frac{\delta L}{\delta \dot{\gamma}_{ij}}\cr
 &=&  \gamma^{ij}K-K^{ij}\
 +\lgb\left[\gamma^{ij}(\mathcal{R}K-2\mathcal{R}_{kl}K^{kl})-\mathcal{R}K^{ij}-2\mathcal{R}^{ij}K+4\mathcal{R}^{k(i}K^{j)}_{\; k}
 \right. \nonumber\\
 &&\hskip 4cm +2\mathcal{R}^{ikjl}K_{kl}+\frac{1}{3}\gamma^{ij}(-K^3+3KK_{kl}^2-2K_k^lK_l^mK_m^k) \nonumber\\
 &&\hskip 5cm \left. +K^2K^{ij}-2KK_k^iK^{jk} -K^{ij}K_{kl}^2+2K_k^iK_l^kK^{jl}\right], \nonumber\\
 \pi_\phi & \equiv& \frac{1}{\sqrt{-\gamma}}\frac{\delta L}{\delta \dot{\phi}} = -\dot{\phi}, \qquad
 \pi_{\chi}  \equiv \frac{1}{\sqrt{-\gamma}}\frac{\delta L}{\delta \dot{\chi}} =-Z(\phi)\dot{\chi} \,.
\eea
In our solution it is clear that $\pi_\chi=0$, but we keep this term in this section for full generality. The Hamiltonian for radial evolution is then given by
\begin{equation}
 H = \int_{\Sigma_r}d^{d-1}x\sqrt{-\gamma}(2\pi^{ij}K_{ij}+\pi_\phi\dot{\phi} + \pi_\chi\dot{\chi}) - L\,,
\end{equation}
where we used that $K_{ij}=\dot{\gamma}_{ij}/2$ in the chosen gauge. To write the Hamiltonian in terms of the canonical momenta and induced metric one needs to invert (\ref{eq:canonicalmomentum}), which is a complicated system of nonlinear equations. This has been done in \cite{LiuSabra}, but only to first order in $\lambda_\mt{GB}$. In this section we also limit ourselves to this regime, for simplicity (although, we repeat, our solution (\ref{ansatz-sol}) is fully non-perturbative in $\lgb$). Using the results from that paper and adding the axion-dilaton contribution, we find that
\bea
 H&=&-\int_{\Sigma_r}d^{d-1}x\sqrt{-\gamma}\Bigl[\mathcal{R}+(d-1)(d-2)-\frac{1}{d-2}(\pi_i^i)^2+\pi_{ij}^2+\pi_\phi^2+\frac{\pi_\chi^2}{Z(\phi)}  \nonumber\\
 &&\hskip 1.5cm-\frac{1}{2}\bigl((\partial_i\phi)^2+Z(\phi)(\partial_i\chi)^2\bigr)+\frac{\lgb}{2}\Bigl(\mathcal{R}^2-4\mathcal{R}_{ij}^2
 +\mathcal{R}_{ijkl}^2-\frac{16}{d-2}\pi_k^k\mathcal{R}_{ij}\pi^{ij}  \nonumber\\
 &&\hskip 1.5cm +\frac{2d}{(d-2)^2}(\pi_i^i)^2\mathcal{R}-2\mathcal{R}\pi_{ij}^2+8\mathcal{R}_{ij}\pi^{jk}\pi^i_k+4\mathcal{R}_{ijkl}\pi^{ik}\pi^{jl}+2\pi^j_i\pi^k_j\pi^l_k\pi^i_l-(\pi_{ij}^2)^2  \nonumber\\
 &&\hskip 1.5cm -\frac{16}{3(d-2)}\pi^l_l\pi_i^j\pi_j^k\pi_k^i+\frac{2d}{(d-2)^2}(\pi_k^k)^2\pi_{ij}^2-\frac{3d-4}{3(d-2)^3}(\pi_i^i)^4\Bigr) \Bigr] +O(\lgb^2)\,.
 \label{Hradev1}
\eea
More details about the derivation of this result are reported in App.~\ref{appB}.


\subsection{Recursive method}

Consider now a regularized space $\mathcal{M}_r$, whose boundary is $\Sigma_r$, with a fixed $r$ which in the end is meant to be taken to infinity. We add a generic boundary term $S_b$ to the action defined on this regularized space. In \cite{Yiannis} it was shown that the variational problem is well defined if
\begin{equation}
 S_b\big|_r=-\mathcal{S}_r,
\end{equation}
where $\mathcal{S}_r$ is Hamilton's principal functional, given by the on-shell action with arbitrary boundary values for $\gamma_{ij},\phi$, and $\chi$ on $\Sigma_r$.

It is well known that the canonical momenta can be obtained by taking functional derivatives of ${\cal S}_r$
\begin{equation} \label{eq:momentum}
 \pi^{ij}=\frac{\delta\mathcal{S}_r}{\delta \gamma_{ij}}, \qquad
 \pi_\phi=\frac{\delta\mathcal{S}_r}{\delta \phi}, \qquad
 \pi_\chi=\frac{\delta\mathcal{S}_r}{\delta \chi}.
\end{equation}
The Hamiltonian is constrained to vanish as a result of the equations of motion for $N$ and $N_i$
\begin{equation}
 H=0\,.
\end{equation}
We can determine $\mathcal{S}_r$ by solving this constraint. The trick is to consider an expansion in eigenfunctions of the operator
\begin{equation}
 \delta_\gamma=\int_{\Sigma_r} d^{d-1}x\,2\gamma_{ij}\vard{}{\gamma_{ij}}.
\end{equation}
One can verify that such an expansion is a derivative expansion
\begin{equation}
 \mathcal{S}_r=\mathcal{S}_{(0)}+\mathcal{S}_{(2)}+\mathcal{S}_{(4)}+\ldots\, ,
\end{equation}
with
\be
\delta_\gamma \mathcal{S}_{(2n)}=(d-1-2n)\mathcal{S}_{(2n)}\,.
\label{eigen}
\ee 
Once we know the solution for $\mathcal{S}_{(0)}$, we can compute corrections to the action in a systematic way by solving algebraic equations. In fact, having to deal with algebraic equations instead of partial differential equations is the main advantage of the method of \cite{Yiannis}. 

Now we write Hamilton's principal functional as
\begin{equation}
 \mathcal{S}_r=\int_{\Sigma_r}d^{d-1}x\,\left(\mathcal{L}_{(0)}+\mathcal{L}_{(2)}+\mathcal{L}_{(4)}+\ldots\right)\,.
\end{equation}
From (\ref{eq:momentum}) we see that the canonical momenta  also admit derivative expansions
\begin{align}
 \pi^{ij} &= {{\pi_{(0)}}}^{ij}+{{\pi_{(2)}}}^{ij}+{\pi_{(4)}}^{ij}+\ldots\,,\cr 
 \pi_\phi&= {\pi_\phi}_{(0)}+{\pi_\phi}_{(2)}+{\pi_\phi}_{(4)}+\ldots\,,\cr 
 \pi_\chi&= {\pi_\chi}_{(0)}+{\pi_\chi}_{(2)}+{\pi_\chi}_{(4)}+\ldots\,.
\end{align}
Translating (\ref{eigen}) in terms of the momenta and the Lagrangian density we obtain
\begin{equation} \label{eq:piL}
 2\pi_{(2n)}=\frac{1}{\sqrt{-\gamma}}(d-1-2n)\mathcal{L}_{(2n)}\,,
\end{equation}
where $\pi_{(2n)}={\pi_{(2n)}}^i_{\;i}$ is the trace taken with respect to $\gamma^{ij}$. This relation is crucial for the algorithm to work. Note that  we can obtain all canonical momenta at some given order by just knowing the trace of the momentum conjugate to the induced metric. 

We can now solve the Hamiltonian constraint $H=0$. Substituting the above expansions in the Hamiltonian and grouping terms with the same number of derivatives leads to an equation of the form
\begin{equation}
  H=H_{(0)}+H_{(2)}+H_{(4)}+\ldots=0\,,
\end{equation}
which must be satisfied order by order, imposing separately $H_{(2n)}=0$ for all $n$.


\subsubsection{Solution at zeroth order}

We start by collecting terms with zero derivatives, which results in
\bea
H_{(0)} & =& -(d-2)(d-1) - {\pi^2_{(0)}}_{ij} +  \frac{\pi_{(0)}^2}{d-2} - {\pi^2_\phi}_{(0)}- \frac{{\pi^2_\chi}_{(0)}}{Z(\phi)} \cr
&&+ \lgb \left( \frac{1}{2} ({\pi^2_{(0)}}_{ij})^2 
-\frac{d\, {\pi^2_{(0)}}_{ij} \pi_{(0)}^2}{(d-2)^2} + \frac{(3d-4) \pi_{(0)}^4}{6(d-2)^3} - {\pi_{(0)}}_{i}^{k} \pi_{(0)}^{ij} \left({\pi_{(0)}}_{j}^{l} {\pi_{(0)}}_{kl} + \frac{8 {\pi_{(0)}}_{jk} \pi_{(0)}}{6 - 3 d}\right)\right)\cr && +O(\lgb^2)\,.
\eea
Following \cite{Yiannis}, we try with the Ansatz for $\mathcal{L}_{(0)}$
\begin{equation}
  \mathcal{S}_{(0)}=2\int_{\Sigma_r}d^{d-1}x\sqrt{-\gamma}\,{\cal W}(\phi,\chi)\,,
 \end{equation}
and compute the corresponding canonical momenta
\begin{equation}
 \pi_{(0)}^{ij}=\vard{\cs_{(0)}}{\gamma_{ij}}=\gamma^{ij}{\cal W}\,,\qquad
 {\pi_\phi}_{(0)}=\vard{\cs_{(0)}}{\phi}=2\pa_\phi {\cal W}\,,\qquad
 {\pi_\chi}_{(0)}=\vard{\cs_{(0)}}{\chi}=2\pa_\chi {\cal W}\,.
\end{equation}
Substituting into $H_{(0)}$ this gives
\bea \label{eqW}
H_{(0)}&=&
- (d-2)(d-1)+\frac{(d-1) {\cal W}^2}{d-2}-\frac{4(\pa_\chi {\cal W})^2}{Z(\phi)}-4(\pa_\phi {\cal W})^2\cr
&& \hskip 3cm+\lgb\left(\frac{(d-4)(d-3)(d-1) {\cal W}^4}{6 (d-2)^3}\right)+O(\lgb^2)\,.
\eea
We know from \cite{Yiannis} that in the limit $\lgb\to 0$ the solution for ${\cal W}$ is the constant $d-2$. We can then write
\begin{equation} \label{W}
 {\cal W}(\phi,\chi) = (d-2)+\lgb {\cal V}(\phi,\chi)+O(\lgb^2).
\end{equation}
Plugging (\ref{W}) into (\ref{eqW}) gives an equation for ${\cal V}(\phi,\chi)$, whose solution turns out to be also a constant
\begin{equation}
 {\cal V}=-\frac{1}{12}(d-4)(d-3)(d-2).
\end{equation}


\subsubsection{Solution at second order}

Terms with two derivatives can be collected into the following expression
\bea
H_{(2)} & =&- {\cal R} - 2 \pi_{(0)}^{ij} {\pi_{(2)}}_{ij} +  \frac{2 \pi_{(0)} \pi_{(2)}}{d-2}
- 2 {\pi_\phi}_{(0)} {\pi_\phi}_{(2)} - \frac{2 {\pi_\chi}_{(0)} {\pi_\chi}_{(2)}}{Z(\phi)} +  \frac{1}{2} (\pa_{i}\phi)^2 +  \frac{1}{2} Z(\phi) (\pa_{i}\chi)^2 \nonumber \\ 
&& - \lgb \left[4  {\cal R}^{ij} {\pi_{(0)}}_{i}^{k} {\pi_{(0)}}_{jk} + 2  {\cal R}_{ikjl} \pi_{(0)}^{ij} \pi_{(0)}^{kl} + \frac{d  \, {\cal R} \pi_{(0)}^2}{(d-2)^2}
+ \frac{2 d \, \pi_{(0)}^{ij} \pi_{(0)}^2 {\pi_{(2)}}_{ij}}{\bigl(d-2\bigr)^2}  \right.\nonumber \\
&& -  \frac{8 {\pi_{(0)}}_{i}^{k} \pi_{(0)}^{ij} \pi_{(0)} {\pi_{(2)}}_{jk}}{d-2} + 4 {\pi_{(0)}}_{i}^{k} \pi_{(0)}^{ij} {\pi_{(0)}}_{j}^{l} {\pi_{(2)}}_{kl} + \frac{8 {\pi_{(0)}}_{i}^{k} \pi_{(0)}^{ij} {\pi_{(0)}}_{jk} \pi_{(2)}}{6 - 3 d}
  + \frac{8 \pi_{(0)}^3 \pi_{(2)}}{3 \bigl(d-2\bigr)^3}
\nonumber \\ 
&&\left. -  \frac{2 d\, \pi_{(0)}^3 \pi_{(2)}}{\bigl(d-2\bigr)^3} - {\pi_{(0)}}_{ij} \left( \frac{8  {\cal R}^{ij} \pi_{(0)}}{d-2}
 + \pi_{(0)}^{ij} \left({\cal R} + 2 \pi_{(0)}^{kl} {\pi_{(2)}}_{kl} - \frac{2 d\, \pi_{(0)} \pi_{(2)}}{(d-2)^2}\right)\right)\right]+O(\lgb^2)\,.\cr&&
\eea
Substitution of the zeroth order solution in $H_{(2)}$ leads to the following simple algebraic equation for $\pi_{(2)}$ 
\begin{equation}
\left(-2+(1 + (d-4) (d-3) \lgb) {\cal R}- \frac{1}{2} (d-4) (d-3) \lgb\right) \pi_{(2)} -  \frac{1}{2} (\pa_{i}\phi)^2-\frac{1}{2} Z(\phi) (\pa_{i}\chi)^2=0\,.
\end{equation}
Solving the above equation and using (\ref{eq:piL}), we obtain $\mathcal{L}_{(2)}$
\bea
\mathcal{L}_{(2)} &=&\frac{\sqrt{-\gamma}}{2(d-3)} \left(2{\cal R} - (\pa_{i}\phi)^2- Z(\phi) (\pa_{i}\chi)^2 +  \lgb\frac{12 - 7 d + d^2}{4}\left(6 {\cal R} + (\pa_{i}\phi)^2
 + Z(\phi) (\pa_{i}\chi)^2\right)\right)\cr&& +O(\lgb^2)\,.
\eea
From this we can compute the momenta at second order
\bea
\pi_{(2)}^{ij} & =&-\frac{1}{8(d-3)}\Bigl[4(2 {\cal R}^{ij}  -  \pa^{i}\phi \pa^{j}\phi -Z(\phi) \pa^{i}\chi \pa^{j}\chi)-2\gamma^{ij} ( 2 {\cal R}-  (\pa_{k}\phi)^2-  Z(\phi)( \pa_{k}\chi)^2)  \nonumber \\ 
&& + \lgb (12 - 7 d + d^2) \left(6 {\cal R}^{ij}  + \pa^{i}\phi \pa^{j}\phi + Z(\phi) \pa^{i}\chi \pa^{j}\chi -  \tfrac{1}{2}\gamma^{ij}( 6{\cal R}+ (\pa_{k}\phi)^2 + Z(\phi) (\pa_{k}\chi)^2)\right)\Bigr] \cr
&&+O(\lgb^2)\,,
\nonumber \\ 
{\pi_\phi}_{(2)} &=&\frac{2 - \tfrac{1}{2} (12 - 7 d + d^2) \lgb}{4 (d-3)}\left(2 D_i D^i\phi - Z'(\phi) (\pa_{i}\chi)^2 \right) +O(\lgb^2)\,, \nonumber \\ 
{\pi_\chi}_{(2)} &=&\frac{2 - \tfrac{1}{2} (12 - 7 d + d^2) \lgb}{2 (d-3)}\left(Z(\phi)D_i D^i\chi +Z'(\phi) \pa_{i}\chi \pa^{i}\phi \right)+O(\lgb^2)\,.
\eea


\subsubsection{Solution at fourth order}

Finally, at fourth order we have
\begin{align}
H_{(4)}& = 
-{\pi_{(2)}}_{ij} \pi_{(2)}^{ij} +  \frac{\pi_{(2)}^2}{d-2} - {\pi^2_\phi}_{(2)} - \frac{{\pi^2_\chi}_{(2)}}{Z(\phi)} - 2 \pi_{(0)}^{ij} {\pi_{(4)}}_{ij} +  \frac{2 \pi_{(0)} \pi_{(4)}}{d-2} - 2 {\pi_\phi}_{(0)} {\pi_\phi}_{(4)}  - \frac{2 {\pi_\chi}_{(0)} {\pi_\chi}_{(4)}}{Z(\phi)}
\cr &- \lgb
 \left[-2 {\cal R}_{ij} {\cal R}^{ij} + \tfrac{1}{2} {\cal R}^2 
+ \tfrac{1}{2} ({\cal R}_{ijkl})^2 - 2 {\cal R} \pi_{(0)}^{ij} {\pi_{(2)}}_{ij} -  \frac{8 {\cal R}^{ij} \pi_{(0)} {\pi_{(2)}}_{ij}}{d-2} + 8 {\cal R}^{ij} {\pi_{(0)}}_{i}{}^{k} {\pi_{(2)}}_{jk} \right.
 \nonumber \\ 
&  + \frac{d \pi_{(0)}^2 {\pi_{(2)}}_{ij} \pi_{(2)}^{ij}}{(d-2)^2}-  \frac{8 \pi_{(0)}^{ij} \pi_{(0)} {\pi_{(2)}}_{i}{}^{k} {\pi_{(2)}}_{jk}}{d-2} + 2 \pi_{(0)}^{ij} \pi_{(0)}^{kl} {\pi_{(2)}}_{ik} {\pi_{(2)}}_{jl} - 2 \pi_{(0)}^{ij} \pi_{(0)}^{kl} {\pi_{(2)}}_{ij} {\pi_{(2)}}_{kl}
\nonumber \\ 
& + 4 {\pi_{(0)}}_{i}{}^{k} \pi_{(0)}^{ij} {\pi_{(2)}}_{j}{}^{l} {\pi_{(2)}}_{kl}  + 4 {\cal R}_{ikjl} \pi_{(0)}^{ij} \pi_{(2)}^{kl} -  {\pi_{(0)}}_{ij} \pi_{(0)}^{ij} {\pi_{(2)}}_{kl} \pi_{(2)}^{kl}
 -  \frac{8 {\cal R}^{ij} {\pi_{(0)}}_{ij} \pi_{(2)}}{d-2} + \frac{2 d {\cal R} \pi_{(0)} \pi_{(2)}}{(d-2)^2} \nonumber \\ 
& + \frac{4 d \pi_{(0)}^{ij} \pi_{(0)} {\pi_{(2)}}_{ij} \pi_{(2)}}{(d-2)^2} -  \frac{8 {\pi_{(0)}}_{i}{}^{k} \pi_{(0)}^{ij} {\pi_{(2)}}_{jk} \pi_{(2)}}{d-2} + \frac{d {\pi_{(0)}}_{ij} \pi_{(0)}^{ij} \pi_{(2)}^2}{(d-2)^2} + \frac{4 \pi_{(0)}^2 \pi_{(2)}^2}{(d-2)^3} -  \frac{3 d \pi_{(0)}^2 \pi_{(2)}^2}{(d-2)^3} \nonumber \\ 
& + \frac{2 d \pi_{(0)}^{ij} \pi_{(0)}^2 {\pi_{(4)}}_{ij}}{(d-2)^2} -  \frac{8 {\pi_{(0)}}_{i}^{k} \pi_{(0)}^{ij} \pi_{(0)} {\pi_{(4)}}_{jk}}{d-2} + 4 {\pi_{(0)}}_{i}^{k} \pi_{(0)}^{ij} {\pi_{(0)}}_{j}^{l} {\pi_{(4)}}_{kl} - 2 {\pi_{(0)}}_{ij} \pi_{(0)}^{ij} \pi_{(0)}^{kl} {\pi_{(4)}}_{kl} \nonumber \\ 
&\left. + \frac{8 {\pi_{(0)}}_{i}^{k} \pi_{(0)}^{ij} {\pi_{(0)}}_{jk} \pi_{(4)}}{6 - 3 d} 
 + \frac{2 d {\pi_{(0)}}_{ij} \pi_{(0)}^{ij} \pi_{(0)} \pi_{(4)}}{(d-2)^2} + \frac{8 \pi_{(0)}^3 \pi_{(4)}}{3 (d-2)^3} -  \frac{2 d \pi_{(0)}^3 \pi_{(4)}}{(d-2)^3}\right] +O(\lgb^2)
\,.
\end{align}
Now we repeat the previous steps. We substitute the zeroth and second order solutions in $H_{(4)}$, which leads to an algebraic equation for $\pi_{(4)}$ that can be readily solved. There is a subtlety due to the fact that the relation (\ref{eq:piL}) is ill defined for $d=5$, which is our case of interest. In fact 
\begin{equation}
 \mathcal{L}_{(4)}=\sqrt{-\gamma}\frac{2}{d-5}\pi_{(4)}.
\end{equation}
However, the Hamilton-Jacobi method can still be applied (see the discussion in \cite{Yiannis}) if we set the radial cut-off to be
\begin{equation}
 r=\frac{1}{d-5}\,,
\end{equation}
and define $\tilde{\mathcal{L}}_{(4)}$ such that $\mathcal{L}_{(4)}|_r=-2r\tilde{\mathcal{L}}_{(4)}|_r$, namely $\tilde{\mathcal{L}}_{(4)}=-\sqrt{-\gamma}\pi_{(4)}$. Proceeding in this way we finally arrive at 
\begin{align}
\frac{96}{r\sqrt{-\gamma}}\mathcal{L}_{(4)}& = -8 {\cal R}\left({\cal R} - (\pa_{i}\phi)^2 - Z(\phi) \pa_{i}\chi \pa^{i}\chi\right) + 4(\pa_{i}\phi)^4 + 12 Z(\phi)(\pa_{i}\chi \pa^{i}\phi)^2 \nonumber \\ \nonumber
& - 4 Z(\phi) (\pa_{i}\phi)^2 (\pa_{j}\chi)^2 + 4 Z(\phi)^2 (\pa_{i}\chi)^4 + 24 {\cal R}_{ij}  \left({\cal R}^{ij} - \pa^{i}\phi \pa^{j}\phi - Z(\phi) \pa^{i}\chi \pa^{j}\chi\right)\\ \nonumber
& +24\left(D_i D^i\phi-\tfrac{Z'(\phi)}{2}(\pa_i\chi)^2\right)^2+24Z(\phi)\left(D_i D^i \chi+\tfrac{Z'(\phi)}{Z(\phi)}(\pa_i\chi\pa^i\phi)\right)^2\\ \nonumber
&+ \lgb  \Big[76 {\cal R}^2 + 48 {\cal R}_{ijkl} {\cal R}^{ijkl}  - 12 {\cal R} \left((\pa_{i}\phi)^2 + Z(\phi) (\pa_{i}\chi)^2\right)  + 2 (\pa_{i}\phi)^4 + 6 Z(\phi) (\pa_{i}\chi \pa^{i}\phi)^2\\ \nonumber
&  - 2 Z(\phi) (\pa_{i}\phi)^2 (\pa_{j}\chi)^2 + 2 Z(\phi)^2 (\pa_{i}\chi)^4 -12 {\cal R}_{ij} \left(23 {\cal R}^{ij} - 3 \pa^{i}\phi \pa^{j}\phi - 3 Z(\phi) \pa^{i}\chi \pa^{j}\chi\right)\\
&\left.    -36\left(D_i D^i\phi-\tfrac{Z'(\phi)}{2}(\pa_i\chi)^2\right)^2-36Z(\phi)\left(D_i D^i \chi+\tfrac{Z'(\phi)}{Z(\phi)}\pa_i\chi\pa^i\phi\right)^2\right] +O(\lgb^2)\,.
\label{L4anom}
\end{align}
Up to some overall factor, this expression coincides with the conformal anomaly, as we shall see in a moment.


\subsection{Fefferman-Graham expansions} 

From the counterterms obtained using the Hamilton-Jacobi method we see that the canonical momenta take the form
\begin{align}
 \pi^{ij} & = \pi_{(0)}^{ij}+\pi_{(2)}^{ij}-2r{}\tilde{\pi}_{(4)}^{ij}+\pi_{(4)}^{ij}+\ldots\,,\cr
 \pi_\phi & = {\pi_\phi}_{(0)}+{\pi_\phi}_{(2)}-2r{{}\tilde\pi_\phi}_{(4)}+{\pi_\phi}_{(4)}+\ldots\,,\cr
 \pi_\chi & = {\pi_\chi}_{(0)}+{\pi_\chi}_{(2)}-2r{{}\tilde\pi_\chi}_{(4)}+{\pi_\chi}_{(4)}+\ldots\,.
\end{align}
The fourth order terms $\pi_{(4)}^{ij}, {\pi_\phi}_{(4)},{\pi_\chi}_{(4)}$ contain the information about the renormalized one-point functions. In order to determine these terms, we proceed with the asymptotic analysis. In Fefferman-Graham (FG) coordinates, the metric reads
\be
ds^2=\ell_{AdS}^2\left(\frac{dv^2}{v^2}+\gamma_{ij}(x,v)\,dx^i dx^j\right)\,.
\label{FGgeneric}
\ee
As already mentioned in Sec.~\ref{sec2}, the AdS radius $\ell_{AdS}$ is given by 
\be
\ell_{AdS}= \frac{1}{\sqrt{F_\mt{bdry}}}\,,
\ee
and $v= e^{-r/\ell_{AdS}}$. Generically, the fields will have the following near-boundary expansions in these coordinates
\bea
\gamma_{ij}&&=\frac{1}{v^2}\Big(g_{(0)ij}+v^2 g_{(2)ij}+v^4 \left(g_{(4)ij}+2 h_{(4)ij}\log v \right)+O(v^6) \Big)\,,\cr
\phi &&=\phi_{(0)}+v^2\phi_{(2)}+v^4\left( \phi_{(4)} + 2\, \tilde{\phi}_{(4)}\log v \right)+O(v^6)\,,\cr
\chi &&=\chi_{(0)}+v^2\chi_{(2)}+v^4\left( \chi_{(4)} + 2\, \tilde{\chi}_{(4)}\log v \right)+O(v^6)\,.
\eea
The coefficients ${g_{(0)}}_{ij},\phi_{(0)}$ and $\chi_{(0)}$ remain undetermined from this analysis, but the other coefficients can be obtained as functions of ${g_{(0)}}_{ij},\phi_{(0)}$ and $\chi_{(0)}$ by substituting the above expansions in (\ref{eq:canonicalmomentum}) and comparing order by order in $v$. For example, comparing terms at order $O(v^2)$ we obtain
\bea
 {g_{(2)}}_{ij} & =&-\frac{1-\lgb}{2}R_{ij}+\frac{1}{4}(1+\lgb)\left(\partial_i\phi_{(0)}\partial_j\phi_{(0)}+Z(\phi_{(0)})\partial_i\chi_{(0)}\partial_j\chi_{(0)}\right)\cr
 && +\frac{1}{24}{g_{(0)}}_{ij}\left(2(1-\lgb)R-(1+\lgb)(\partial_k\phi_{(0)}\partial^k\phi_{(0)}+Z(\phi_{(0)})\partial_k\chi_{(0)}\partial^k\chi_{(0)})\right)\cr
 && +O(\lgb^2)\,, \cr
 \phi_{(2)} &=&\frac{1-\lgb}{8}\left(2 D_{(0)i}\partial^i\phi_{(0)}-Z'(\phi_{(0)})\pa_i\chi_{(0)}\pa^i\chi_{(0)}\right)+O(\lgb^2)\,,\cr
 \chi_{(2)} &=&\frac{1-\lgb}{4 Z(\phi_{(0)})}\left(Z(\phi_{(0)})D_{(0)i}\partial^i\chi_{(0)}+Z'(\phi_{(0)})\pa_i\chi_{(0)}\pa^i\phi_{(0)}\right)+O(\lgb^2)\,.
\eea
Here and in the following the curvatures $R$ and $R_{ij}$ are the ones for $g_{(0)ij}$. Comparing the logarithmic term, we obtain instead
\bea
\tilde{\pi}_{(4)ij} & =&2\left(1 -\lgb\right)\left( {h_{(4)}}_{ij} - {h_{(4)}}^{k}{}_{k} {g_{(0)}}_{ij}\right)+O(\lgb^2) \,,\cr
\tilde{\phi}_{(4)} & =&\frac{1}{4}(1-\lgb) {\tilde{\pi}}_{\phi(4)}+O(\lgb^2) \,,\qquad
\tilde{\chi}_{(4)}  = \frac{ 1-\lgb}{4 Z(\phi_{(0)})}{\tilde{\pi}}_{\chi(4)} +O(\lgb^2)\,.
\eea


\subsection{The 1-point functions}

The order $O(v^4)$ leads to the following renormalized one-point functions, which represent the main result of our analysis in this section. For the stress tensor we get
\bea
 \vev{ T_{ij}} &=&2{\pi_{(4)}}_{ij} \nonumber \\
 &=&-2 {g_{(2)}}_{ij} {g_{(2)}}^{k}{}_{k} + 4 {g_{(4)}}_{ij} + 2 {h_{(4)}}_{ij} + 2 {g_{(2)}}_{kl} {g_{(2)}}^{kl} {g_{(0)}}_{ij} - 4 {g_{(4)}}^{k}{}_{k} {g_{(0)}}_{ij} - 2 {h_{(4)}}^{k}{}_{k} {g_{(0)}}_{ij}\nonumber \\
 &&+  \tfrac{1}{2} {g_{(2)}}^{kl} {g_{(0)}}_{ij} R_{kl} - \tfrac{1}{2} {g_{(2)}}_{ij} R  - \tfrac{1}{2} {D_{(0)}}_{i}\phi_{(2)} {D_{(0)}}_{j}\phi_{(0)} - \tfrac{1}{2} {D_{(0)}}_{i}\phi_{(0)} {D_{(0)}}_{j}\phi_{(2)}
 \cr &&
  - \tfrac{ Z(\phi_{(0)})}{2} {D_{(0)}}_{i}\chi_{(2)} {D_{(0)}}_{j}\chi_{(0)} +  \tfrac{Z(\phi_{(0)})}{2} {g_{(0)}}_{ij}  {D_{(0)}}_{k}\chi_{(2)} {D_{(0)}}^{k}\chi_{(0)}\nonumber \\ 
&& - \tfrac{ Z(\phi_{(0)})}{2} {D_{(0)}}_{i}\chi_{(0)} {D_{(0)}}_{j}\chi_{(2)} - \tfrac{1}{2} {D_{(0)}}_{j}{D_{(0)}}_{i}{g_{(2)}}^{k}{}_{k} +  \tfrac{1}{2} {D_{(0)}}_{k}{D_{(0)}}_{i}{g_{(2)}}_{j}{}^{k} \nonumber \\ 
&& +  \tfrac{1}{2} {D_{(0)}}_{k}{D_{(0)}}_{j}{g_{(2)}}_{i}{}^{k} - \tfrac{1}{2} D_{(0)k}D^k_{(0)}{g_{(2)}}_{ij} +  \tfrac{1}{4} {g_{(2)}}_{ij} {D_{(0)}}_{k}\phi_{(0)} {D_{(0)}}^{k}\phi_{(0)} \nonumber \\ 
&& +  \tfrac{1}{2} {g_{(0)}}_{ij} {D_{(0)}}_{k}\phi_{(2)} {D_{(0)}}^{k}\phi_{(0)} +  \tfrac{Z(\phi_{(0)})}{4} {g_{(2)}}_{ij}  {D_{(0)}}_{k}\chi_{(0)} {D_{(0)}}^{k}\chi_{(0)}  \nonumber \\ 
&& - \tfrac{1}{2} {g_{(0)}}_{ij} {D_{(0)}}_{l}{D_{(0)}}_{k}{g_{(2)}}^{kl} +  \tfrac{1}{2} {g_{(0)}}_{ij} D_{(0)\ell}D^\ell_{(0)}{g_{(2)}}^{k}{}_{k} - \tfrac{1}{4} {g_{(2)}}_{kl} {g_{(0)}}_{ij} {D_{(0)}}^{k}\phi_{(0)} {D_{(0)}}^{l}\phi_{(0)} \nonumber \\ 
&& - \tfrac{Z(\phi_{(0)})}{4} {g_{(2)}}_{kl} {g_{(0)}}_{ij}  {D_{(0)}}^{k}\chi_{(0)} {D_{(0)}}^{l}\chi_{(0)} - \tfrac{Z'(\phi_{(0)})}{2} \phi_{(2)} {D_{(0)}}_{i}\chi_{(0)} {D_{(0)}}_{j}\chi_{(0)}  \nonumber \\ 
&& +  \tfrac{Z'(\phi_{(0)})}{4} {g_{(0)}}_{ij} \phi_{(2)} {D_{(0)}}_{k}\chi_{(0)} {D_{(0)}}^{k}\chi_{(0)}  +\lgb {T_\mt{GB}}_{ij}+O(\lgb^2)\,,
\label{stress-final}
\eea
where
\bea
{T_\mt{GB}}_{ij}& =&
-4 {g_{(2)}}_{i}{}^{k} {g_{(2)}}_{jk} + 7 {g_{(2)}}_{ij} {g_{(2)}}^{k}{}_{k} - 6 {g_{(4)}}_{ij} - 3 {h_{(4)}}_{ij} - {g_{(2)}}_{kl} {g_{(2)}}^{kl} {g_{(0)}}_{ij} 
\cr &&
- 2 ({g_{(2)}}^{k}{}_{k})^2 {g_{(0)}}_{ij} + 6 {g_{(4)}}^{k}{}_{k} {g_{(0)}}_{ij} + 3 {h_{(4)}}^{k}{}_{k} {g_{(0)}}_{ij} 
+  \tfrac{13}{4} {g_{(2)}}_{ij} R - 2 {g_{(2)}}^{k}{}_{k} {g_{(0)}}_{ij} R
\nonumber \\ 
&&  +\tfrac{29}{2} {g_{(2)}}^{kl} R_{ikjl}
 + 4 {g_{(2)}}^{k}{}_{k} R_{ij} - \tfrac{53}{4} {g_{(2)}}_{j}{}^{k} R_{ik} - \tfrac{53}{4} {g_{(2)}}_{i}{}^{k} R_{jk} +  \tfrac{11}{4} {g_{(2)}}^{kl} {g_{(0)}}_{ij} R_{kl}  \nonumber \\ 
&& +  \tfrac{1}{4} {D_{(0)}}_{i}\phi_{(2)} {D_{(0)}}_{j}\phi_{(0)} +  \tfrac{1}{4} {D_{(0)}}_{i}\phi_{(0)} {D_{(0)}}_{j}\phi_{(2)} +  \tfrac{Z(\phi_{(0)})}{4}  {D_{(0)}}_{i}\chi_{(2)} {D_{(0)}}_{j}\chi_{(0)} \nonumber \\ 
&& +  \tfrac{Z(\phi_{(0)})}{4}  {D_{(0)}}_{i}\chi_{(0)} {D_{(0)}}_{j}\chi_{(2)} +  \tfrac{37}{4} {D_{(0)}}_{j}{D_{(0)}}_{i}{g_{(2)}}^{k}{}_{k} - \tfrac{37}{4} {D_{(0)}}_{i}{D_{(0)}}_{k}{g_{(2)}}_{j}^{k} \nonumber \\ 
&& - \tfrac{37}{4} {D_{(0)}}_{j}{D_{(0)}}_{k}{g_{(2)}}_{i}^{k} +  \tfrac{37}{4} D_{(0)k}D^k_{(0)}{g_{(2)}}_{ij} - \tfrac{1}{8} {g_{(2)}}_{ij} {D_{(0)}}_{k}\phi_{(0)} {D_{(0)}}^{k}\phi_{(0)} \nonumber \\ 
&& - \tfrac{1}{4} {g_{(0)}}_{ij} {D_{(0)}}_{k}\phi_{(2)} {D_{(0)}}^{k}\phi_{(0)} - \tfrac{Z(\phi_{(0)}) }{8} {g_{(2)}}_{ij} {D_{(0)}}_{k}\chi_{(0)} {D_{(0)}}^{k}\chi_{(0)}\cr &&
 - \tfrac{Z(\phi_{(0)})}{4} {g_{(0)}}_{ij}  {D_{(0)}}_{k}\chi_{(2)} {D_{(0)}}^{k}\chi_{(0)} 
  - \tfrac{Z'(\phi_{(0)})}{8} {g_{(0)}}_{ij} \phi_{(2)} {D_{(0)}}_{k}\chi_{(0)} {D_{(0)}}^{k}\chi_{(0)} \nonumber \\ 
&& +  \tfrac{5}{4} {g_{(0)}}_{ij} {D_{(0)}}_{l}{D_{(0)}}_{k}{g_{(2)}}^{kl} - \tfrac{5}{4} {g_{(0)}}_{ij} D_{(0)\ell}D^\ell_{(0)}{g_{(2)}}^{k}{}_{k} +  \tfrac{1}{8} {g_{(2)}}_{kl} {g_{(0)}}_{ij} {D_{(0)}}^{k}\phi_{(0)} {D_{(0)}}^{l}\phi_{(0)} \nonumber \\ 
&& +  \tfrac{Z(\phi_{(0)})}{8} {g_{(2)}}_{kl} {g_{(0)}}_{ij}  {D_{(0)}}^{k}\chi_{(0)} {D_{(0)}}^{l}\chi_{(0)} +  \tfrac{Z'(\phi_{(0)})}{4} \phi_{(2)} {D_{(0)}}_{i}\chi_{(0)} {D_{(0)}}_{j}\chi_{(0)} \,.\cr&&
\eea
For the dilaton and axion we get instead
\bea
\vev{\mathcal{O}_\phi}&=&- {\pi_\phi}_{(4)} \nonumber\\
&= &- (2+\lgb)(2 \phi_{(4)} + \tilde{\phi}_{(4)}) 
\cr &&
+\tfrac{1}{4} (2-\lgb)\left[D_{(0)i}D^i_{(0)}\phi_{(2)} +  
\tfrac{1}{2} {D_{(0)}}_{i}{g_{(2)}}^{j}{}_{j} {D_{(0)}}^{i}\phi_{(0)} - {D_{(0)}}^{i}\phi_{(0)} {D_{(0)}}_{j}{g_{(2)}}_{i}{}^{j}
\right. \nonumber \\ 
&&\hskip 2.5cm 
- {g_{(2)}}_{ij} {D_{(0)}}^{j}{D_{(0)}}^{i}\phi_{(0)}
 +  \tfrac{ Z'(\phi_{(0)})}{2} {g_{(2)}}_{ij} {D_{(0)}}^{i}\chi_{(0)} {D_{(0)}}^{j}\chi_{(0)} \nonumber \\ 
&& \hskip2.5cm \left.
- Z'(\phi_{(0)}) {D_{(0)}}_{i}\chi_{(2)} {D_{(0)}}^{i}\chi_{(0)}   - \tfrac{Z''(\phi_{(0)})}{2} \phi_{(2)} {D_{(0)}}_{i}\chi_{(0)} {D_{(0)}}^{i}\chi_{(0)} \right]
\cr&& +O(\lgb^2)\,,\cr&&
\eea
and
\bea
  \vev{\mathcal{O}_\chi}&=&- {\pi_\chi}_{(4)} \nonumber\\
  &=&
- (2+\lgb)\left(2 Z(\phi_{(0)}) \chi_{(4)} 
 +Z(\phi_{(0)})\tilde{\chi}_{(4)} +Z'(\phi_{(0)}) \phi_{(2)} \chi_{(2)}\right)
 \cr&&+ 
\tfrac{1}{4} (2-\lgb)
 \left[Z(\phi_{(0)})  D_{(0)i}D^i_{(0)}\chi_{(2)} +  \tfrac{Z(\phi_{(0)}) }{2} {D_{(0)}}_{i}{g_{(2)}}^{j}{}_{j} {D_{(0)}}^{i}\chi_{(0)} \right.\nonumber \\ 
&&  \hskip2.5cm -Z(\phi_{(0)})  {D_{(0)}}^{i}\chi_{(0)} {D_{(0)}}_{j}{g_{(2)}}_{i}{}^{j} -
Z(\phi_{(0)}) {g_{(2)}}_{ij}  {D_{(0)}}^{j}{D_{(0)}}^{i}\chi_{(0)}  \nonumber \\ 
&&  \hskip2.5cm+  Z'(\phi_{(0)}) \phi_{(2)} D_{(0)i}D^i_{(0)}\chi_{(0)} + Z'(\phi_{(0)}) {D_{(0)}}_{i}\chi_{(2)} {D_{(0)}}^{i}\phi_{(0)}  \nonumber \\ 
&&  \hskip2.5cm+ Z'(\phi_{(0)}) {D_{(0)}}_{i}\chi_{(0)} {D_{(0)}}^{i}\phi_{(2)}  - Z'(\phi_{(0)}) {g_{(2)}}_{ij} {D_{(0)}}^{i}\phi_{(0)} {D_{(0)}}^{j}\chi_{(0)}  \nonumber \\ 
&&\left. \hskip2.5cm + Z''(\phi_{(0)}) \phi_{(2)} {D_{(0)}}_{i}\chi_{(0)} {D_{(0)}}^{i}\phi_{(0)}\right]\cr&&
+O(\lgb^2)\,.\cr &&
\label{chi-final}
\eea
We stress that these formulas are generic for any axion-dilaton system with GB term (to first order in $\lgb$) of the structure given in (\ref{total-action}). 

The zeroth order terms in $\lgb$ in these expressions reproduce the results of \cite{Yiannis}, while the first order terms in $\lgb$ extend the results of \cite{LiuSabra} to a system with an axion-dilaton field. As mentioned already, another difference with the analysis of \cite{LiuSabra} is that we employ a recursive method which is more effective in cases where multiple fields, besides the metric, are turned on. 


\subsection{Central charges}

The trace of the stress energy tensor is related to the central charges $a$ and $c$ by the following expression\footnote{Notice that in this section $a$ denotes one of the central charges and not the anisotropy parameter.}
\be
 \vev{T^i_i}=\frac{1}{16\pi^2} \left( c\, W - a\, E \right)+\ldots\,,
\label{anomaly1}
\ee
where $E$ is the four-dimensional Euler density
\begin{equation}
E={\cal R}^2-4 {\cal R}_{ij}{\cal R}^{ij}+{\cal R}_{ijkl}{\cal R}^{ijkl}\,,
\end{equation}
$W$ is the square of the Weyl tensor
\be
W=C^{ijkl} C_{ijkl} =\frac{{\cal R}^2}{3}-2 {\cal R}_{ij}{\cal R}^{ij}+{\cal R}_{ijkl}{\cal R}^{ijkl}\,,
\label{W2}
\ee
and where the ellipsis indicates the contribution by other fields (the axion-dilaton in our specific setting). The trace of the stress energy tensor  \cite{Yiannis} is given by the ${\cal L}_{(4)}$ written above in (\ref{L4anom})
\begin{align}
\vev{T_i^i}= \frac{1}{r\sqrt{-\gamma}}\mathcal{L}_{(4)}\,.
\end{align}
To isolate the metric contribution we set $\phi = \chi = 0$ in that expression and arrive at
\begin{align}
\vev{T^i_i}& = -\frac{1}{12} {\cal R}^2
 + \frac{1}{4} {\cal R}_{ij}  {\cal R}^{ij}+ \left(\frac{19}{24} {\cal R}^2 + \frac{1}{2} {\cal R}_{ijkl} {\cal R}^{ijkl}   -\frac{23}{8} {\cal R}_{ij}  {\cal R}^{ij} \right)\lgb
 +O(\lgb^2)\,.
\label{anomaly2}
\end{align}
Comparing (\ref{anomaly1}) and (\ref{anomaly2}), we find that  
\be
a=\pi^2(2-15\, \lgb)+O(\lgb^2)\,,\qquad c=\pi^2(2-7\, \lgb)+O(\lgb^2)\,,
\ee
thus confirming that indeed $a\neq c$ for theories with GB corrections. These results are in perfect agreement with previous literature, see e.g.  \cite{Hung:2011xb}.


\section{Boundary stress tensor}
\label{sec4}

Here we specialize the formulas above to our solution (\ref{ansatz-sol}). As a first step, we need to rewrite the fields in FG coordinates, to be able to extract the asymptotic behaviors close to the boundary. 

We define the FG radial coordinate $v$ such that\footnote{Asymptotically, our metric approaches $AdS_5$ with curvature radius given by $\ell_{AdS}=1/\sqrt{F_\mt{bdry}}$, which explains the factor of $F_\mt{bdry}$ in the formula, see (\ref{FGgeneric}).}
\be
\frac{du^2}{u^2 F(u)}=\frac{dv^2}{v^2 F_\mt{bdry}}+O(v^3)\,,\qquad F_\mt{bdry}=\frac{1-\sqrt{1-4\lgb}}{2\lgb}\,.
\ee
The relation between the two radial coordinates $u$ and $v$ turns out to be given explicitly by
\bea
u&=&v+\frac{1}{48} a^2 (\lgb +1) v^3\cr
&&\hskip 1.5cm -\frac{a^2 \uh^2 (2 \lgb  (\log 32-1)+1+\log 4)+12 (\lgb +1)}{96 \uh^4}v^5+O(a^4,\lgb^2,v^7)\,.\cr&&
\eea
In terms of $v$ the fields have the following asymptotic expansions
\bea
\phi(v)&=&-\frac{a^2}{4}(1-\lgb)v^2+\frac{a^2}{8\uh^2}\left(1-\lgb\right)v^4+O(v^6)\,,\cr
F(v)&=&1+\lgb+\frac{a^2}{12}\left(1+2\lgb\right)v^2\cr &&
\hskip 1.5cm-\left(\frac{1+2\lgb}{\uh^4}+\frac{a^2}{12 \uh^2}\left(1+2\log 2-(1-12\log 2)\lgb\right)\right)v^4+O(v^6)\,,\cr
B(v)&=& 1-\lgb -\frac{a^2}{12}v^2+\frac{a^2}{8\uh^2}v^4+O(v^6)\,,\cr
H(v)&=&1+\frac{a^2}{4}(1+\lgb)v^2-\frac{a^2}{8\uh^2}(1+\lgb)v^4+O(v^6)\,.
\eea
From these expressions it is easy to find the expansions for the metric
\bea
g_{tt}&=&-1+\frac{a^2}{24}(1+\lgb)v^2\cr 
&&\hskip .5cm
 +\frac{1}{16\uh^4}\left(12(1+\lgb)-a^2\uh^2(1-2\log 2+2\lgb(2-5\log 2))\right)v^4+O(v^6)\,,\cr
g_{xx}&=&g_{yy}=1-\frac{a^2}{24}(1+\lgb)v^2
\cr &&\hskip .5cm
+\frac{1}{48\uh^4}\left(12(1+\lgb)+a^2\uh^2(1+2\log 2-2\lgb(1-5\log 2))\right)v^4+O(v^6)\,,\cr
g_{zz}&=&1+\frac{5a^2}{24}(1+\lgb)v^2
\cr &&\hskip .5cm
+\frac{1}{48\uh^4}\left(12(1+\lgb)-a^2\uh^2(5-2\log 2+2\lgb(4-5\log 2))\right)v^4+O(v^6)\,,\cr&&
\eea
from which it is immediate to extract $g_{(2)ij}$ and $g_{(4)ij}$. 

In our solution  the terms up to  $O(a^2)$ are very simple:
\bea
\vev{T_{ij}} = 4 {g_{(4)}}_{ij} -  6\lgb {g_{(4)}}_{ij}\,, \qquad {\pi_\phi}_{(4)}  = 0 \,,\qquad 
{\pi_\chi}_{(4)} =0\,.
\label{useful4}
\eea
Explicitly, the components of the stress tensor read
\bea
\vev{T_{tt}} &=&\frac{3}{\uh^4}-\frac{1-2\log 2}{4 \uh^2}a^2-\frac{12+a^2 \uh^2 (5-14 \log 2)}{8 \uh^4}\lgb
   + O(a^4,\lgb^2)\,,\cr
\vev{T_{xx}} & =&\vev{T_{yy}}=\frac{1}{\uh^4}+\frac{1+2\log 2}{12 \uh^2}a^2-\frac{12+a^2 \uh^2 (7-14 \log 2)}{24 \uh^4}\lgb+O(a^4,\lgb^2)\,, \cr
\vev{ T_{zz}} &=&\frac{1}{\uh^4}-\frac{5-2\log 2}{12 \uh^2}a^2-\frac{12+a^2 \uh^2 (1-14 \log 2)}{24 \uh^4}\lgb+O(a^4,\lgb^2)\,.
\eea
Using (\ref{temppos}) we see that
\be
\uh=\frac{1}{\pi T}-\frac{1-\log 2}{24\pi^3T^3}a^2-\frac{1}{2\pi T}\left(1-\frac{5\log 2}{24\pi^2T^2}a^2\right)\lgb+O(a^4,\lgb^2)\,,\ee
so that we can rewrite the expressions above in terms of the temperature, which is a physical observable, unlike  the horizon location $\uh$. We arrive at our final results:
\bea
\vev{T_{tt}} &=&3\pi^4T^4\left[1+\frac{1}{12\pi^2}\left(\frac{a}{T}\right)^2+\left(\frac{3}{2}+\frac{1}{24\pi^2}\left(\frac{a}{T}\right)^2\right)\lgb\right]
   + O(a^4,\lgb^2)\,,\cr
\vev{T_{xx}} & =&\vev{T_{yy}}=\pi^4T^4\left[1+\frac{1}{4\pi^2}\left(\frac{a}{T}\right)^2+\left(\frac{3}{2}+\frac{1}{8\pi^2}\left(\frac{a}{T}\right)^2\right)\lgb\right]
+O(a^4,\lgb^2)\,, \cr
\vev{ T_{zz}}&=&\pi^4T^4\left[1-\frac{1}{4\pi^2}\left(\frac{a}{T}\right)^2+\left(\frac{3}{2}-\frac{1}{8\pi^2}\left(\frac{a}{T}\right)^2\right)\lgb\right]
+O(a^4,\lgb^2)\,.
\label{stress2}
\eea

These quantities correspond to the energy density and pressures of the dual gauge theory
\be
E=\frac{\nc^2}{8\pi^2}\vev{T_{tt}}\,,\qquad P_\perp=\frac{\nc^2}{8\pi^2}\vev{T_{xx}}\,,\qquad P_\|=\frac{\nc^2}{8\pi^2}\vev{T_{zz}}\,,
\ee
with $\nc$ being the number of colors of the gauge theory and $P_\perp$ and $P_\|$ the pressures along the transverse plane and the longitudinal direction, respectively. The comparison with the energy density $E_0(T)=3\pi^2\nc^2T^4/8$ and the pressure $P_0(T)=\pi^2 \nc^2 T^4/8$ of an isotropic plasma at the same temperature and $\lgb=0$ is obvious from the expressions above. We see in particular that the anisotropy has the effect of increasing the energy density and perpendicular pressure compared to the isotropic case, while decreasing the longitudinal pressure. This is consistent with the findings of \cite{MT,MT2} in the small anisotropy limit (whose results we reproduce for $\lgb=0$, see eq. (168) of \cite{MT2}). 

These results show that the system is really anisotropic in the $z$-direction, as $P_\perp\neq P_\|$. Notice that at this order in $a$, the trace of the stress tensor is vanishing
\be
\vev{T^i_i}=O(a^4,\lgb^2)\,.
\ee
This is in agreement with what found in \cite{MT,MT2}, where the conformal anomaly was also vanishing at order $O(a^2)$ and appearing only at order $O(a^4)$ and beyond. We can also check some basic thermodynamic relations. In particular, the free energy ${\cal F}=E-T s$, in the limit of $a=0$, matches perfectly the value found in \cite{Brigante} from evaluating the Euclidean action on-shell. We can also check that ${\cal F}=-P_\perp$, as it should be \cite{emparan}.

Finally, let us comment about the conservation of the (expectation value of the) stress tensor. Remember that to simplify our expressions we have gauge fixed the lapse and shift functions (see footnote \ref{shiftnote}). As a consequence we can no longer derive the diffeomorphism Ward identity that relates the divergence of $\vev{T_{ij}}$ to the expectation values of the other fields. Typically, in such an identity we expect a term of the form $\vev{{\cal O}_\chi}\partial_j\chi_{(0)}$, see e.g. eq. (B.20) of \cite{Yiannis}. In the particular state we have considered, even though $\partial_j\chi_{(0)}\neq 0$, we do have that $\vev{{\cal O}_\chi}=0$, see (\ref{useful4}). This contribution would then vanish, assuring that $D^i_{(0)}\vev{T_{ij}}=0$ and guaranteeing the translational invariance of the geometry. For a more detailed study of the thermodynamics of this system, which is beyond the scope of the present paper, one would need to derive this Ward identity.


\section{The dual anisotropic plasma}
\label{sec5}

As anticipated in the Introduction, one application of the solution we have found is modeling higher curvature effects on the dual gauge theory plasma. Generically, heavy ion collisions in experiments are non-central, resulting in a spatial anisotropy of the QGP formed in the collision. This represents one of the main motivations for our Ansatz. In the following we will identify with $z$ the anisotropic direction (or `beam' direction), while $x$ and $y$ parametrize the plane transverse to the beam. 

\subsection{Shear viscosity to entropy ratios}

An important quantity to compute in a plasma is the ratio of shear viscosity over entropy density.\footnote{Other observables that have been computed in Einstein plus GB gravity can be found in, e.g., \cite{other,other1}.} This is a rather universal quantity for theories with an Einstein dual, which has been conjectured to obey the Kovtun-Son-Starinets (KSS) bound $\eta/s\ge 1/4\pi$ \cite{KSS}. This bound can however be violated by the inclusion of higher derivative corrections \cite{Brigante} (see also \cite{beyond,beyond0,beyond05,beyond1,beyond2,beyond23,beyond25,beyond3,beyond4,beyond5}) and by the breaking of spatial isotropy \cite{Rebhan,mamo}; see \cite{sera} for a status report on the viscosity bound.

In this section, we employ the {\it membrane paradigm}, proposed in \cite{Iqbal} and used in \cite{Rebhan} for the anisotropic plasma of \cite{MT}, to compute $\eta/s$ for our geometry (\ref{ansatz-sol}).\footnote{The computation of the shear viscosity in an anisotropic superfluid with a GB term has recently been presented in \cite{anis_superfluid}.} Appendix \ref{app2} contains two alternative derivations of the results in this section.\footnote{Yet another way of doing the computation would be the so-called Riccati equation method, developed in \cite{riccati1} and revisited in \cite{riccati2}. This method allows to obtain the 2-point functions directly from the canonical momenta of Sec.~\ref{sec3}, without deriving any effective action.} As in \cite{Rebhan}, we will be interested in two components of the shear viscosity tensor: $\eta_{xyxy}$, which is entirely in the transverse (isotropic) plane, and $\eta_{xzxz}=\eta_{yzyz}$, which mixes the anisotropic direction $z$ with one of the directions in the transverse plane. We denote these two components as
\be
\eta_{\perp}=\eta_{xyxy}\,,\qquad  \eta_\| = \eta_{xzxz}\,.
\ee 
To calculate these viscosities we consider the fluctuations $h_{xy}$ and $h_{xz}$ around the background (\ref{ansatz-sol}). Given the symmetry in the transverse plane, we can take these fluctuations to depend solely on $(t,y,z,u)$. The equations of motion for $\psi_{\perp} =h^{x}_{\,\,y}(t,y,z,u)$ and $\psi_\| =h^{x}_{\,\,z}(t,y,z,u)$ decouple from all other equations and from each other. In both cases, they have the following form
\be
a(u) \psi''+b(u)\psi'+c(u)\psi=0\,,
\label{eqpsi}
\ee
where $a(u), b(u)$ and $c(u)$ are functions of the background fields and $\psi$ stands for either $\psi_{\perp}$ or $\psi_{\|}$, depending on the case. Here the primes denote derivatives with respect to $u$. To use the membrane paradigm, we need to write an effective action for $\psi_{\perp}$ and $\psi_{\|}$. To this scope we write (\ref{eqpsi}) in the form\footnote{It is important to emphasize that $n(u)$ and $m(u)$ are not the same in the equations of motion for $\psi_{\perp}$ and $\psi_{\|}$. Here $n(u)$ stands for either $n_{\perp}$ or $n_{\|}$, and $m(u)$ stands for either $m_{\perp}$ or $m_{\|}$.} 
\be
(n(u) \psi')'-m(u)\psi=0\,,
\ee
with
\be
n(u)=\mathrm{exp} \left(\int _u du' \frac{a(u')}{b(u')} \right)\,,\qquad
m(u)=\frac{c(u)}{a(u)} \mathrm{exp} \left( \int_u du' \frac{a(u')}{b(u)} \right)\,.
\ee
The effective action that gives rise to the equation of motion above is
\be
S_\mt{eff} = -\frac{1}{2}\frac{1}{16\pi G} \int d^4x \, du \left[n(u)(\psi')^2-m(u)\psi^2   \right]\,.
\label{action-eff}
\ee
To compare this action with the one of \cite{Iqbal}, we need to transform it to Fourier space. To do that, we write
\be
\psi(t,y,z,u) = \int \frac{d \omega}{2 \pi} \frac{d^3k}{(2 \pi)^3} \psi(u) e^{-i\omega t + i k_y y + i k_z z}\,,
\label{psiF}
\ee
where we have used the axial symmetry to rotate ${\bf k}=(0, k_y, k_z)$. Plugging (\ref{psiF}) into (\ref{action-eff}) and using Plancherel's theorem, it can be shown that
\be
S_\mt{eff} = -\frac{1}{2} \frac{1}{16\pi G} \int \frac{d \omega}{2 \pi} \frac{d^3k}{(2 \pi)^3} du \left[n(u)(\psi')^2-m(u)\psi^2   \right]\,.
\label{Seff}
\ee
Using the notation of \cite{Iqbal}, this can be recast in the following form
\be
S_\mt{eff}= -\frac{1}{2} \int \frac{d \omega}{2 \pi} \frac{d^3k}{(2 \pi)^3} du \sqrt{-g} \left[\frac{g^{uu}}{Q(u,k)}(\psi')^2+P(u,k) \psi^2   \right]\,,
\label{Smb}
\ee
with
\be
\frac{1}{16\pi G} n(u)= \frac{\sqrt{-g}\,g^{uu}}{Q(u,k)} \,.
\label{eqQ}
\ee
The shear viscosity is then obtained as \cite{Iqbal}
\be
\frac{\eta}{s} = \frac{1}{4 \pi} \frac{16\pi G}{Q(u_{\mt{H}},k \rightarrow 0)} \,.
\label{eqeta}
\ee
Writing the equations of motion for  $\psi_{\perp}$ and $\psi_{\|}$, we can obtain explicit expressions for the $n(u)$'s and $m(u)$'s. Putting these together with (\ref{eqQ}) and (\ref{eqeta}), it is readily found that
\bea
\frac{\eta_{\perp}}{s}&=&\frac{1}{4 \pi}\left(\frac{g_{xx}}{g_{yy}}-\frac{\lgb}{2}\frac{g_{xx}^2 g_{tt}'g_{zz}'}{g}  \right)\,,\cr
\frac{\eta_{\|}}{s} &=&\frac{1}{4 \pi}\left(\frac{g_{xx}}{g_{zz}}-\frac{\lgb}{2}\frac{g_{xx}^2 g_{tt}'g_{yy}'}{g}  \right)\,.
\label{etageneric}
\eea
These results are completely generic for the system we have considered. In particular, we can check them against the known results from pure Einstein-Hilbert gravity with a GB term
\cite{Brigante} and with the anisotropic background of \cite{Rebhan}, finding perfect agreement in both cases. In the first case, we need to take the limit of $a \rightarrow 0$ of the equations above. We find 
\be
\frac{\eta_{\perp}}{s}=\frac{\eta_{\|}}{s} = \frac{1-4 \lgb}{4 \pi}\,,
\ee
as in \cite{Brigante}. To perform the second check we take the limit $\lgb \rightarrow 0$ and obtain\footnote{Note that to compare the expressions for $\eta_\|$ one needs to take into account the different factors of the dilaton in the Ans\"azte of \cite{MT,Rebhan} and (\ref{ansatz-metric}).}
\be
\frac{\eta_{\perp}}{s}=\frac{1}{4 \pi}\,, \qquad \frac{\eta_{\|}}{s}=\frac{1}{4 \pi}\frac{1}{H(u_{\mt{H}})}
=\frac{1}{4\pi}-\frac{\log 2}{16\pi^3}\left(\frac{a}{T}\right)^2+O(a^4)\,.
\ee
Note how the longitudinal shear viscosity violates the KSS bound.

Specializing (\ref{etageneric}) to our solution (\ref{ansatz-sol}) we find
\bea
\frac{\eta_{\perp}}{s}& =&\frac{1-4 \lgb }{4 \pi }+\frac{B_0}{24\pi^3}\frac{ \lgb  (3-4 \lgb )}{ ( 1-4 \lgb)}\left(\frac{a}{T}\right)^2+O\left(a^4\right)\,,\cr
\frac{\eta_{\|}}{s}&=&\frac{1-4\lgb}{4 \pi}+\frac{B_0}{32 \pi^3} G(\lgb)\left(\frac{a}{T}\right)^2+O(a^4)\,,
\label{shear-result}
\eea
where $G(\lgb)$ is given by
\bea
G(\lgb)& = &-1+2 \lgb  \left(\frac{8 \lgb }{12 \lgb -3}+1\right)+\sqrt{1-4 \lgb }\nonumber \\
&& \hspace{1cm} +\sqrt{\lgb } \log \left(\frac{1+2 \sqrt{\lgb  }}{1-2 \sqrt{\lgb }}\right)+\log\left(\frac{\sqrt{1-4 \lgb}-1+4 \lgb }{8 \lgb }\right)\,.
\eea
We emphasize that these results, despite being of second order in $a$, are fully nonperturbative in $\lgb$. The KSS bound might be violated in this setting both by the anisotropy and by the GB coupling.


\subsection{Conductivities}

To calculate the plasma conductivities,\footnote{For a related computation in an isotropic background with linear scalar fields and a GB term see \cite{chinese_new}.} we need to introduce\footnote{We introduce this field only in this section, solely for the purpose of computing the conductivities. Of course, the analysis of Sec.~\ref{sec3} would be modified by the inclusion of an extra field.} a $U(1)$ gauge field $A_m$ in the bulk, with a standard Maxwell action
\be
S_\mt{Maxwell}=-\int d^5x \sqrt{-g}\frac{1}{4 g_\mt{eff}^2(u)} F_{mn}F^{mn}\,,
\ee
where $g_{\mathrm{eff}}(u)$ is a generic $u$-dependent coupling constant. The conjugate momentum to the gauge field is given by\footnote{In this section we keep denoting the boundary coordinates by the Latin indices $i,j,\ldots$, as in Sec.~\ref{sec3}.}
\be
j^{i}=-\frac{\sqrt{-g}}{g_\mt{eff}^2}F^{i u}\,.
\label{eqj}
\ee
The gauge field $A_{m}$ is dual to a conserved current $J^i$ in the boundary theory whose expectation value is equal to $j^i$ evaluated at the boundary
\be
\vev{J^i(k)} = j^i(u \rightarrow 0\ ; k)\,.
\ee
The AC conductivity is given by the following relation between the spatial part of $j^i$ and the electric field $F_{jt}$
\be
\vev{J^{i=x,y,z}(k)} = \sigma^{ij}(k)F_{jt}(u \rightarrow 0)\,,
\label{sigmadef}
\ee
while the DC conductivity is defined by the zero momentum limit of $\sigma^{ij}(k)$
\be
\sigma^{ij}_\mt{DC}= \lim_{k\rightarrow 0 } \sigma^{ij}(k)\,.
\ee
It turns out that we can calculate these quantities doing a near horizon analysis \cite{Iqbal}. Imposing infalling boundary conditions at the horizon implies that
\be
F_{ui}=\sqrt{-\frac{g_{uu}}{g_{tt}}}F_{ti}\Big|_{u=\uh}\,.
\label{eqreg}
\ee
Combining (\ref{eqj}) and (\ref{eqreg}) it can be readily shown that
\be
j^i(\uh)=\frac{1}{g_\mt{eff}^2}\sqrt{\frac{g}{g_{tt}g_{uu}}}g^{ij}F_{jt}(\uh)\,,
\label{jsigmaE}
\ee
and that, in the zero momentum limit, this relation holds for all $u$ \cite{Iqbal}. Because of this, we can do the calculation at the horizon, instead of doing it at the boundary. Comparing (\ref{sigmadef}) with (\ref{jsigmaE}) we see that the conductivity along the $i$-direction is
\be
\sigma^{ii}_\mt{DC}=\frac{1}{g_\mt{eff}^2}\sqrt{\frac{g}{g_{tt}g_{uu}}}g^{ii}\Big|_{\uh}\,.
\ee
For an isotropic background we have $\sigma^{ij}_\mt{DC} = \sigma \delta^{ij}$ (from now on we are going to use the symbol $\sigma$ to represent the DC conductivity and will drop the subscript). When the background is anisotropic, there will be two different conductivities: $\sigma_{\perp}$ and $\sigma_\|$. The former corresponds to an electric field aligned along the $x$- and $y$-directions, resulting in a corresponding conductivity along the transverse plane, whereas the latter corresponds to electric field and conductivity along the beam direction. These quantities are given by
\bea
\sigma_{\perp}=\frac{1}{g^2_\mt{eff}}\sqrt{\frac{g}{g_{tt}g_{uu}}}g^{xx}\Big|_{u_{\mt{H}}}\,,\qquad 
\sigma_\|=\frac{1}{g^2_\mt{eff}}\sqrt{\frac{g}{g_{tt}g_{uu}}}g^{zz}\Big|_{\uh}=\frac{\sigma_{\perp}}{H(\uh)}\,.
\eea
Normalizing with the isotropic result, we get
\bea
&&\frac{\sigma_{\perp}}{\sigma_\mt{iso}}=H(u_{\mt{H}})^{1/2}=1+\frac{a^2}{2}H_2(\uh)+O(a^4)\,,\cr
&&\frac{\sigma_\|}{\sigma_\mt{iso}}=H(u_{\mt{H}})^{-1/2}=1-\frac{a^2}{2}H_2(\uh)+O(a^4)\,.
\eea
Since $H_2(\uh)$ is a positive quantity, we see that the anisotropy has the effect of enhancing the conductivity along the perpendicular directions, as compared to the isotropic case, while suppressing the one along the longitudinal direction, consistently with the findings of \cite{photon,dileptons}.


\section{Conclusion}
\label{sec6}

In this paper we have explored the effects of higher curvature corrections (given by the inclusion of a GB term) in a system of AdS-gravity in five dimensions coupled to an axion-dilaton field. As we have explained above, these corrections correspond, on the gauge theory side, to considering cases that are more generic than the ones usually considered, e.g. conformal field theories with independent central charges, $a\neq c$. It is still unclear whether our setup might be obtained in the low energy limit of some string theory, and our philosophy has been `bottom-up'. 

One of our main concerns has been to carry out holographic renormalization and compute  the 1-point function of the boundary stress tensor associated to our gravitational theory. We have done this to first order in the GB coupling,  which is however not a terribly restrictive constraint, since requirements of unitarity, causality and positivity of energy fluxes require \cite{reviewlovelock}
\be
-7/36\le \lgb \le 9/100\,.
\label{constrlambda}
\ee 

We have also considered a particular black brane Ansatz, in which the axion field is linearly dependent on one of the horizon coordinates, while being independent of the radial coordinate. This has resulted in finding an anisotropic black brane solution to the equations of motion, which is the GB-corrected equivalent of the geometry discovered in \cite{MT,MT2}. We have computed the shear viscosity over entropy density ratio for the dual plasma and found that the KSS bound \cite{KSS} is violated, as expected from previous works where either the case $(a=0,\lgb\neq 0)$ \cite{Brigante} or the case $(a\neq0,\lgb=0)$ \cite{Rebhan} were considered. As discussed in Sec.~\ref{sec2}, one point that remains to be settled in our analysis is whether our solution might be interpreted as an interpolating solution between a Lifshitz-like scaling solution in the IR and an asymptotically AdS space, as was the case for the $\lgb=0$ limit of \cite{MT,MT2}. 

One of the most interesting applications of the present work would be a detailed study of the thermodynamics of this black brane and of its corresponding plasma. This analysis was carried out, in the canonical ensemble, for the case of vanishing $\lgb$ in \cite{MT,MT2} and a rich phase diagram was discovered, with, in particular, the presence of instabilities that might turn out to be useful in understanding the fast thermalization time of the QGP. To this regard it is relevant to observe that part of the richness of the solution in \cite{MT,MT2} was due to a conformal anomaly, appearing in the renormalization process at order $O(a^4)$ and beyond. In the present solution we also have an anomaly, which we expect to appear at the fourth order in the anisotropy parameter, but we are not able to capture with our analytic solution, which only goes up to second order. Extending our analytic solution to order $O(a^4)$ seems unviable and presumably numerical methods would have to be employed to explore larger values of 
the anisotropy. Given the large number of parameters in the game, this might be cumbersome, but it surely is something worth pursuing. 

Finally, one could study other physical observables besides the shear viscosity and conductivities, such as the energy loss via dragging and quenching, the quark-antiquark screening or the production of thermal photons, to name a few.


\section*{Acknowledgements}
We are happy to thank David Mateos and Ioannis Papadimitriou for insightful comments on the draft. We are supported in part by CNPq and by FAPESP grants 2014/01805-5 (VJ), 2014/07840-7 (ASM), and 2013/02775-0 (DT).


\appendix
 
\section{Derivation of the solution}
\label{app1}

In this Appendix we give some details on how we have found our solution (\ref{ansatz-sol}) and present its explicit expression. 

The Einstein equations (\ref{eom}) are diagonal, as a consequence of the fact that the metric only depends on $u$. We have then four equations for the metric (since the $xx$- and $yy$-components are not independent) plus the equation for the dilaton in (\ref{eom-dil}). There are four fields to solve for: $\phi$, $F$, $B$, and $H$. Plugging the Ansatz (\ref{ansatz-metric})-(\ref{ansatz-sol}) into the equations and expanding to order $O(a^2)$ one finds that the equation for $\phi_2(u)$ decouples. It reads
\be
\phi_2''+\frac{u\, F_0'-3 F_0}{u\, F_0}\phi_2'=\frac{1}{F_0}\,,
\ee
with $F_0$ given by (\ref{F0}). This can be readily solved changing coordinates as
\be
u\to U(u)=\sqrt{1-4\lgb\left(1-\frac{u^4}{\uh^4}\right)}
\label{changecoord}
\ee
in intermediate steps. The two integration constants are fixed in such a way that $\phi_2$ is regular at the horizon and vanishes at the boundary, $\phi_{2,\mt{bdry}}=0$. One finds 
\bea
\phi_2(u)&=&-\frac{\uh^2 }{8} \left[\alpha+U(u)+\log\left(1+\frac{u^2}{\uh^2}\right)^2
\right.\cr
&& \hskip 1cm \left.
-\sqrt\lgb \log\left(U(u)+2\sqrt\lgb\frac{u^2}{\uh^2}\right)^2
-\log\left(U(u)+1-4\lgb\left(1+\frac{u^2}{\uh^2}\right)\right)\right]\,,\cr&&
\label{phi2sol}
\eea
where
\be
\alpha\equiv -\sqrt{1-4\lgb}+\sqrt\lgb\log\left(1-4\lgb\right)+\log\left(1-4\lgb+\sqrt{1-4\lgb}\right)\,,
\ee
and $U(u)$ is defined as above. We notice that $U$ is always positive (since $\lgb<1/4$), and so is the argument of the last logarithm in (\ref{phi2sol}). When $\lgb=0$ we recover the result of \cite{MT2}, see eq. (164) of that paper.

To find $H_2$, we take the difference of the $xx$- and $zz$-components of (\ref{eom}). One obtains a decoupled equation that reads 
\be
H_2''(u) +p(u) H_2'(u)=q(u)\,,
\ee
with
\bea
p(u)&=&\frac{3(1-4\lgb)(U(u)-1)+4\lgb(3U(u)-5)u^4/\uh^4}{u \,U(u)^2\, (1-U(u))}\,,\cr
q(u)&=&\frac{2\lgb U(u)}{(1-4\lgb)(1-U(u))}\,.
\eea
This equation can be integrated readily via (\ref{changecoord}), fixing the integration constants as above. In particular we request that $H_{2,\mt{bdry}}=0$. The final result is
\bea
H_2(u)&=&\frac{\uh^2}{8(1-4\lgb)}\left[\beta+
U(u)+\log\left(1+\frac{u^2}{\uh^2}\right)+2\lgb\frac{u^2}{\uh^2}\left(\frac{u^2}{\uh^2}-2\right)\right.\cr
&& \hskip 1cm \left. -\sqrt\lgb\log\left(U(u)+2\sqrt\lgb\frac{u^2}{\uh^2}\right)^2-\log\left(
\frac{U(u)+1-4\lgb\left(1+\frac{u^2}{\uh^2}\right)}{U(u)-1+4\lgb\left(1+\frac{u^2}{\uh^2}\right)}\right)^{1/2}
\right]\,,\cr&&
\eea
where, again, we have left $U(u)$ implicit in some places for compactness and where
\bea
\beta \equiv-\sqrt{1-4\lgb}+\sqrt\lgb\log(1-4\lgb)+\log\left(\frac{1+\sqrt{1-4\lgb}}{2\sqrt\lgb}\right)\,.
\eea

Similarly we can solve for the other fields. More specifically, now that we know $\phi_2$ and $H_2$, we can use the $tt$-component of (\ref{eom}) to obtain $F_2$ and the $uu$-component to obtain $B_2$. One can finally check that the $xx$- and $zz$-components are also solved separately, as expected because of the Bianchi identities. The explicit expressions for the equations are not particularly illuminating, so that we limit ourselves to reporting the final results for the remaining fields, which are given by
\bea
F_2(u)&=&\frac{\uh^2}{12(1-4\lgb)U(u)}\left(\frac{u}{\uh}\right)^4\left[
\gamma+U(u)+(1-4\lgb)\left(\frac{\uh}{u}\right)^2
\right.\cr &&\hskip1cm
+4\lgb\left(\frac{u}{\uh}\right)^2-6\lgb\left(\frac{u}{\uh}\right)^4+\log\left(1+\frac{u^2}{\uh^2}\right)^2
\cr
&&\hskip 1cm \left.-\sqrt\lgb\log\left(U(u)+2\sqrt\lgb\frac{u^2}{\uh^2}\right)^2
-\log\left(U(u)+1-4\lgb\left(1+\frac{u^2}{\uh^2}\right)\right)\right]\,,\cr&&
\eea
with 
\be
\gamma\equiv -2+6\lgb+\sqrt\lgb\log\left(1+2\sqrt\lgb\right)^2 +\log\left(\frac{1-4\lgb}{2}\right)\,,
\ee
and by
\bea
B_2(u)&=&\frac{\uh^2}{24(1-4\lgb)}\left[\alpha+U(u)\frac{\uh^2-u^2}{\uh^2+u^2}+\log\left(1+\frac{u^2}{\uh^2}\right)^2
\right. \cr && \hskip 1cm 
-\frac{2u^2}{\uh^2+u^2}\left(1-2\lgb+\lgb\left(\frac{u}{\uh}\right)^2+3\lgb\left(\frac{u}{\uh}\right)^4\right)
\cr && \hskip 1cm
\left.
-\sqrt\lgb\log\left(U(u)+2\sqrt\lgb\frac{u^2}{\uh^2}\right)^2
-\log\left(U(u)+1-4\lgb\left(1+\frac{u^2}{\uh^2}\right)\right)
\right]\,. \cr&&
\eea
Again, we have fixed the integration constants in such a way that the fields be regular at the horizon and vanish at the boundary, $F_{2,\mt{bdry}}=B_{2,\mt{bdry}}=0$. Notice also that $F_2(\uh)=0$, as it should be for a blackening factor. One can check that when $\lgb=0$ the results from \cite{MT} are recovered.\footnote{In order to do so, one needs to take into account the different Ans\"atze and include a factor of the dilaton in (\ref{ansatz-metric}), according to eq. (8) of \cite{MT2}.}


\section{Derivation of the Hamiltonian}
\label{appB}

Here we derive the expression for the Hamiltonian of radial evolution in (\ref{Hradev1}).  The starting point is
\begin{equation}
 H = \int_{\Sigma_r}d^{d-1}x\sqrt{-\gamma}(2\pi^{ij}K_{ij}+\pi_\phi\dot{\phi} + \pi_\chi\dot{\chi}) - L\,,
\end{equation}
where we write $L=\tilde{L}+L_\mt{axion-dilaton}$, with
\begin{equation}
 L_\mt{axion-dilaton}=-\frac{1}{2}\int_{\Sigma_r}d^{d-1}x\sqrt{-\gamma}\left[\dot{\phi}^2+Z(\phi)\dot{\chi}^2+(\partial_i\phi)^2+Z(\phi)(\partial_i\chi)^2\right]\,.
\end{equation}
We can then separate
\begin{equation}
 H=
 \underbrace{\left(\int_{\Sigma_r}d^{d-1}x\sqrt{-\gamma}\,2\pi^{ij}K_{ij} - \tilde{L}\right)}_{\equiv\tilde{H}} +
 \underbrace{\left(\int_{\Sigma_r}d^{d-1}x\sqrt{-\gamma}\left(\pi_\phi\dot{\phi} + \pi_\chi\dot{\chi}\right) - L_\mt{axion-dilaton}\right)}_{\equiv H_\mt{axion-dilaton}}.
\end{equation}
We note that $\tilde{H}$ is exactly the Hamiltonian in eq. (2.12) of \cite{LiuSabra} (up to an overall minus sign). For $H_\mt{axion-dilaton}$ we have 
\begin{align}
 H_\mt{axion-dilaton}=-\int_{\Sigma_r}d^{d-1}x\sqrt{-\gamma}\left(\pi_\phi^2+\frac{\pi_\chi^2}{Z(\phi)}-\frac{1}{2}(\partial_i\phi)^2-\frac{1}{2}Z(\phi)(\partial_i\chi)^2\right)\,.
\end{align}
Writing this in terms of the canonical momenta and induced metric leads to (\ref{Hradev1}).


\section{Shear viscosity tensor}
\label{app2}

In this Appendix we report two alternative derivations of the shear viscosity tensor (\ref{shear-result}).

\subsection{Kubo formula}

As is well known (see e.g. \cite{SS,PSS,PSS2,KS}), the shear viscosity can be also computed using a Kubo formula
\begin{equation} \label{eq:kubo}
\eta = \lim_{\omega\to0}\frac{1}{\omega}\text{Im}\,G_\mt{R}(\omega,\vec{k}=0),
\end{equation}
where $G_\mt{R}(k)$ is the retarded Green's function for the stress tensor. First, we take metric fluctuations $h_{mn}$ around our solution and linearize the equations of motion. Here, we are interested in the modes $\psi_\perp=h^{x}_{\;y}$ and $\psi_\|=h^{x}_{\;z}$. In momentum space, we have
\begin{equation}
 \psi(u,x)=\int \frac{d^4k}{(2\pi)^4}J(k)\psi(u;k)e^{-ik_i x^i},\qquad k_i=(-\omega,{\bf k}),
\end{equation}
where $\psi$ denotes generically one of the modes $\psi_\perp$ or $\psi_\|$. The prescription tells us to solve the equation for $\psi(u;k)$ imposing infalling boundary conditions and regularity at the horizon and satisfying $\psi=1$ at the boundary. 

To compute the shear viscosity, we can restrict ourselves to zero spatial momentum and small frequency $\omega$. For simplicity, we also consider small $\lgb$. The linearized equations for $\psi(u;\omega)$ have the form
\begin{equation}
 K_0(u)\psi''+K_0'(u)\psi'=0,
\end{equation}
where for $\psi=\psi_\perp$ we have, up to orders $O(a^4,\lgb^2,\omega^2)$,
\bea
K_0^\perp(u)& =&\frac{u^4 \left(a^2\uh^2 \log 2+6\right)-a^2\uh^6 \log   \left(1+\frac{u^2}{\uh^2}\right)-6\uh^4}{12 u^3 \uh^4}\cr
&&+\frac{\lgb}{12 u^3\uh^8}  
\Big[u^8\left(a^2\uh^2 (5-6\log2)-18\right)-u^4\uh^4 \left(a^2\uh^2 (2-5\log 2)-6\right)
\cr &&\hskip 2cm
-4 a^2 u^6\uh^4+(12+a^2 u^2)\uh^8+a^2\uh^2 \left(3 u^8-2\uh^8\right) \log \left(1+\frac{u^2}{\uh^2}\right)\Big]\,,\cr&&
\eea
and for $\psi=\psi_\|$ we have
\bea
K_0^\|(u)& =&K_0^\perp(u)+\frac{a^2 \left(\uh^4-u^4\right) \log \left(1+\frac{u^2}{\uh^2}\right)}{8 u^3   \uh^2}
\cr &&
+\frac{a^2 \lgb  \left(-7 u^8+10 u^6 \uh^2-u^4 \uh^4-2 u^2   \uh^6+2 \left(3 u^8-5 u^4 \uh^4+2 \uh^8\right) \log\left(1+\frac{u^2}{\uh^2}\right)\right)}{16 u^3 \uh^6}\,.
\cr&&
\eea
The equations above can be solved by considering an Ansatz of the form
\begin{equation}
\psi(u;\omega)=\left(1-\frac{u^4}{\uh^4}\right)^{-\frac{i\omega}{4\pi T}}\Big[1+\omega\left(f_0(u)+\lgb(f_1(u)+a^2f_2(u))\right) + O(a^4,\lgb^2,\omega^2)\Big]\,,
\end{equation}
where $T$ is the temperature given by (\ref{temppos}). The functions $f_0(u),f_1(u)$ and $f_2(u)$ can be determined by substituting the Ansatz into the linearized equation and solving order by order. The resulting expressions are not particularly illuminating and we do not report them here. The next step is to compute the quadratic on-shell action, which turns out to be a surface term of the form
\begin{equation}
 S^{(2)}_\mt{on-shell}=-\frac{1}{2} \int \frac{d^4k}{(2\pi)^4}J(k)\mathcal{F}(u;k)J(-k)\Big|^{u=\uh}_{u\to0},
\end{equation}
with $\mathcal{F}(u,k)=\frac{1}{16\pi G} K_0(u)\,\psi'(u;k)\psi(u;-k)$. The prescription of \cite{SS} instructs us to take only the contribution of the boundary. The retarded Green's function is then given by
 \begin{equation}
 G_\mt{R}(k)=\lim_{u\to0}\mathcal{F}(u;k).
 \end{equation}
Finally, using (\ref{eq:kubo}) and the result for the entropy (\ref{entropy}) we can compute the ratio of the shear viscosity over entropy density
\begin{align}
\frac{\eta_\perp}{s} & =\frac{1-4\lgb}{4\pi}+ a^2\lgb\frac{\uh^2}{8\pi}+O(a^4,\lgb^2)\,,\cr
\frac{\eta_\|}{s} & =\frac{1-4\lgb}{4\pi} + a^2\left(3\lgb- 2\log 2\right)\frac{\uh^2}{32\pi}+O(a^4,\lgb^2)\,.
\end{align}
These results agree with the ones obtained via the membrane paradigm expanded to first order in $\lgb$.


\subsection{Near-horizon matching technique}

In this section we calculate the viscosities $\eta_{\perp}$ and $\eta_{\|}$ using the near-horizon matching technique of \cite{Kiritsis}. We first solve the fluctuation equation for $\omega = 0$ and then expand the solution near the horizon. After that, we reverse the order of the operations, finding first a near-horizon solution and then expanding it for small $\omega$. Matching the two solutions we obtain the retarded correlator $G_\mt{R}$ from which we can calculate $\eta_{\perp}$ and $\eta_{\|}$.

Consider a fluctuation $\psi$ (again, $\psi$ could be $\psi_{\perp} =h^{x}_{\,\,y}(t,y,z,u)$ or $\psi_{||} =h^{x}_{\,\,z}(t,y,z,u)$, as in the main body of this paper). The effective action and the equation of motion for $\psi$ have the following form
\bea
&&S_\mt{eff}= -\frac{1}{2} \frac{1}{16\pi G} \int \frac{d \omega}{2 \pi} \frac{d^3k}{(2 \pi)^3} du \left[n(u)(\psi')^2-m(u)\psi^2   \right]\,,\cr
&&(n(u) \psi')'-m(u)\psi=0\,,\qquad  \psi = \psi(u,k)\,.
\eea
To be concrete, let us work out the case $\psi = \psi_{\|}$. For ${\bf k}=0$, we have
\bea
n_{\|}(u)&=&g^{uu}\sqrt{-g}\left(\frac{g_{xx}}{g_{zz}}-\frac{\lgb}{2}\frac{g_{xx}^2 g_{tt}'g_{yy}'}{g}  \right)\,,\cr
m_{\|}(u)&=&-\omega^2 g^{tt}\sqrt{-g}\frac{g_{xx}}{g_{zz}}-\frac{\lgb}{2}\omega^2 \left[\frac{g_{yy} g_{uu}' g_{yy}'+g_{yy}'^2-2 g_{yy} g_{yy}'}{\sqrt{g_{tt} g_{zz}}(g_{uu} g_{yy})^{3/2}} \right] g_{xx}^{3/2}\,.
\eea
When $\omega = 0$ and ${\bf k}=0$, we get that $m_{\|}(u)=0$ and $(n(u) \psi')'=0$. Hence $n(u) \psi'= C_2$, where $C_2$ is a constant. This implies that
\be
\psi_{\|}=C_1+C_2 \int_{0}^{u}\frac{du'}{n_{\|}(u')}\,.
\label{sol1}
\ee
As $n_{\|} \propto g^{uu}$ and $g^{uu}$ goes to zero at the horizon, we must have $C_2=0$ for $\omega = 0$. For small $\omega$ we can have a normalizable solution with $C_2 \propto \omega$. Using the prescription of \cite{SS} we calculate $G_\mt{R}$ as
\be
G_\mt{R}=-2\left[-\frac{1}{2}\frac{1}{16\pi G}n_{\|}(u) \psi_{\|}(u,-k)\partial_u \psi_{\|}(u,k) \right]_{u=0}\,,
\label{G}
\ee
where $\psi_{\|}$ should be equal to one at the boundary $u=0$. Using (\ref{sol1}) and (\ref{G}) we can see that
\be
G_\mt{R}=\frac{C_2}{16\pi G}\,,\qquad  (\mathrm{small}\,\,\omega, \, {\bf k}=0)\,.
\label{GR}
\ee
We set $C_1=1$ in the equation above in order to have $\psi_{\|}(0,k)=1$. Now we have to determine $C_2$. Near the horizon, (\ref{sol1}) can be written as
\be
\psi_{\|}=1+C_2\int_0^{u}\frac{du'}{n_{\|}'(u_{\mt{H}})(u-u_{\mt{H}})}=1+\frac{C_2}{n_{\|}'(u_{\mt{H}})}\mathrm{log}\left(1-\frac{u}{u_{\mt{H}}}\right)\,.
\label{solpsi1}
\ee
We now find a near-horizon expression for $\psi_{\|}$ and expand it for small $\omega$. The first thing we need to do is to write the $\psi_{\|}$ equation of motion in the limit $u \rightarrow u_{\mt{H}}$. In what follows, it will be convenient to work with the constants $c_0$ and $c_1$, defined by the near-horizon expansions for $g_{tt}$ and $g^{uu}$ as
\bea
g_{tt}=c_0 (u-u_{\mt{H}})\,,\qquad
g^{uu}=c_1 (u-u_{\mt{H}})\,.
\eea
The near-horizon equation of motion is
\be
(n_{\|}(u_{\mt{H}})(u-u_{\mt{H}}) \psi_{\|}')'-m_{\|}(u_{\mt{H}})\psi_{\|}=0\,,
\label{eqh}
\ee
where
\bea
&&n_{\|}(u_{\mt{H}})= c_1 \sqrt{-g}\left(\frac{g_{xx}}{g_{zz}}-\frac{\lgb}{2}\frac{g_{xx}^2 g_{tt}'g_{yy}'}{-g} \right)\Big|_{u_{\mt{H}}}\,,\cr
&&m_{\|}(u_{\mt{H}})=-\frac{\omega^2}{c_0 (u-u_{\mt{H}})}\sqrt{-g}\,\frac{g_{xx}}{g_{zz}}\Big|_{u_{\mt{H}}} \,.
\label{nmh}
\eea
Plugging (\ref{nmh}) into (\ref{eqh}) and using the anzats $\psi_{\|}=(1-u/u_{\mt{H}})^\beta$ we can show that $\beta =\pm i \omega /\sqrt{c_0 c_1}$.\footnote{Note that the temperature is given by $T = \frac{\sqrt{c_0 c_1}}{4 \pi}$, so that $\beta =\pm \frac{i \omega}{4 \pi T}$.} The general solution of (\ref{eqh}) is then given by
\be
\psi_{\|}=C_+ \left(1-\frac{u}{u_{\mt{H}}} \right)^{\frac{i\omega}{\sqrt{c_0 c_1}}}+C_- \left(1-\frac{u}{u_{\mt{H}}} \right)^{-\frac{i\omega}{\sqrt{c_0 c_1}}}\,.
\ee
We choose $C_+ = 0$ to impose infalling boundary conditions at the horizon. For small $\beta$, i.e. for $\omega \ll T$, we have
\be
\psi_{\|}=C_{-}\left[ 1+\frac{i \omega}{\sqrt{c_0 c_1}}\mathrm{log}\left(1-\frac{u}{u_{\mt{H}}}\right)\right]\,.
\label{solpsi2}
\ee
Comparing (\ref{solpsi2}) with (\ref{solpsi1}) we can see that
\be
\frac{C_2}{n_{\|}(u_{\mt{H}})}=\frac{-i \omega}{\sqrt{c_0 c_1}}\,,
\ee
and using (\ref{nmh}) it is easy to show that
\be
C_2 = -i \omega \sqrt{\frac{c_1}{c_0}}\sqrt{-g}\left(\frac{g_{xx}}{g_{zz}}-\frac{\lgb}{2}\frac{g_{xx}^2 g_{tt}'g_{yy}'}{-g} \right)\Big|_{u_{\mt{H}}}\,.
\ee
Plugging this result into (\ref{GR}) we have
\be
G_\mt{R} = -\frac{i \omega}{16\pi G}\sqrt{\frac{c_1}{c_0}}\sqrt{-g}\left(\frac{g_{xx}}{g_{zz}}-\frac{\lgb}{2}\frac{g_{xx}^2 g_{tt}'g_{yy}'}{-g} \right)\Big|_{u_{\mt{H}}}\,.
\ee
According to the Kubo formula
\be
\eta_{\|}=\lim_{\omega \rightarrow 0} \frac{1}{\omega} \mathrm{Im}\,G_\mt{R}
=\frac{1}{16\pi G}\sqrt{\frac{c_1}{c_0}}\sqrt{-g}\left(\frac{g_{xx}}{g_{zz}}-\frac{\lgb}{2}\frac{g_{xx}^2 g_{tt}'g_{yy}'}{-g} \right)\Big|_{u_{\mt{H}}}\,.
\ee
The density entropy $s$ is given by
\be
s =\frac{1}{4G}\sqrt{\frac{g}{g_{tt}g_{uu}}}\Big|_{u_{\mt{H}}}= 4 \pi \frac{\sqrt{-g}}{16\pi G}\sqrt{\frac{c_1}{c_0}}\,,
\ee
and we can finally calculate the ratio
\be
\frac{\eta_{\|}}{s} =\frac{1}{4 \pi}\left(\frac{g_{xx}}{g_{zz}}-\frac{\lgb}{2}\frac{g_{xx}^2 g_{tt}'g_{yy}'}{g}  \right)\Big|_{u_{\mt{H}}}\,.
\ee
Doing the same for $\psi_{\perp}$ it is possible to show that
\be
\frac{\eta_{\perp}}{s}=\frac{1}{4 \pi}\left(\frac{g_{xx}}{g_{yy}}-\frac{\lgb}{2}\frac{g_{xx}^2 g_{tt}'g_{zz}'}{g}  \right)\Big|_{u_{\mt{H}}}\,.
\ee
In both cases we find the same result (\ref{shear-result}) that we had obtained using the membrane paradigm.



\begin{thebibliography}{99}

\bibitem{Maldacena:1997re}
J.~M.~Maldacena,
``The large $N$ limit of superconformal field theories and supergravity,''
Adv.\ Theor.\ Math.\ Phys.\ {\bf 2}, 231 (1998)
[Int.\ J.\ Theor.\ Phys.\ {\bf 38}, 1113 (1999)]
[hep-th/9711200].

\bibitem{duality2}
  S.~S.~Gubser, I.~R.~Klebanov, A.~M.~Polyakov,
  ``Gauge theory correlators from noncritical string theory,''
  Phys.\ Lett.\  {\bf B428}, 105-114 (1998) [hep-th/9802109].
 
 \bibitem{duality3}
  E.~Witten,
  ``Anti-de Sitter space and holography,''
  Adv.\ Theor.\ Math.\ Phys.\  {\bf 2}, 253-291 (1998) [hep-th/9802150].

\bibitem{lovelock1} 
  D.~Lovelock,
  ``The Einstein tensor and its generalizations,''
  J.\ Math.\ Phys.\  {\bf 12}, 498 (1971).
  
 \bibitem{lovelock2}
  B.~Zwiebach,
  ``Curvature Squared Terms and String Theories,''
  Phys.\ Lett.\ B {\bf 156}, 315 (1985).
  
\bibitem{lovelock3}  
 D.~G.~Boulware and S.~Deser,
  ``String Generated Gravity Models,''
  Phys.\ Rev.\ Lett.\  {\bf 55}, 2656 (1985).
  
\bibitem{lovelock4}  
   R.~-G.~Cai,
  ``Gauss-Bonnet black holes in AdS spaces,''
  Phys.\ Rev.\ D {\bf 65}, 084014 (2002)
  [hep-th/0109133].
  
  \bibitem{reviewlovelock0} 
  T.~Padmanabhan and D.~Kothawala,
  ``Lanczos-Lovelock models of gravity,''
  Phys.\ Rept.\  {\bf 531}, 115 (2013)
  [arXiv:1302.2151 [gr-qc]].
  
\bibitem{reviewlovelock} 
  J.~D.~Edelstein,
  ``Lovelock theory, black holes and holography,''
 Springer Proc.\ Math.\ Stat.\  {\bf 60}, 19 (2014)
  [arXiv:1303.6213 [gr-qc]].
  
  \bibitem{reviewlovelock2}
  X.~O.~Camanho, J.~D.~Edelstein and J.~M.~Sanchez De Santos,
  ``Lovelock theory and the AdS/CFT correspondence,''
  Gen.\ Rel.\ Grav.\  {\bf 46}, 1637 (2014)
  [arXiv:1309.6483 [hep-th]].
  
   \bibitem{branches} 
  X.~O.~Camanho and J.~D.~Edelstein,
  ``Causality in AdS/CFT and Lovelock theory,''
  JHEP {\bf 1006}, 099 (2010)
  [arXiv:0912.1944 [hep-th]].
  
  \bibitem{camanhoetal} 
  X.~O.~Camanho, J.~D.~Edelstein, G.~Giribet and A.~Gomberoff,
  ``Generalized phase transitions in Lovelock gravity,''
 Phys.\ Rev.\ D {\bf 90}, 064028 (2014)
  [arXiv:1311.6768 [hep-th]].
  
    \bibitem{bestiary} 
  X.~O.~Camanho and J.~D.~Edelstein,
  ``A Lovelock black hole bestiary,''
  Class.\  Quant.\  Grav.\  {\bf 30}, 035009 (2013)
  [arXiv:1103.3669 [hep-th]].
  
    \bibitem{ALT} 
  T.~Azeyanagi, W.~Li and T.~Takayanagi,
  ``On String Theory Duals of Lifshitz-like Fixed Points,''
  JHEP {\bf 0906}, 084 (2009)
  [arXiv:0905.0688 [hep-th]].

 \bibitem{MT} 
  D.~Mateos and D.~Trancanelli,
  ``The anisotropic N=4 super Yang-Mills plasma and its instabilities,''
  Phys.\ Rev.\ Lett.\  {\bf 107}, 101601 (2011)
  [arXiv:1105.3472 [hep-th]].

\bibitem{MT2}
  D.~Mateos and D.~Trancanelli,
  ``Thermodynamics and Instabilities of a Strongly Coupled Anisotropic Plasma,''
  JHEP {\bf 1107}, 054 (2011)
  [arXiv:1106.1637 [hep-th]].

\bibitem{Yiannis} 
  I.~Papadimitriou,
  ``Holographic Renormalization of general dilaton-axion gravity,''
  JHEP {\bf 1108}, 119 (2011)
  [arXiv:1106.4826 [hep-th]].

\bibitem{LiuSabra} 
  J.~T.~Liu and W.~A.~Sabra,
  ``Hamilton-Jacobi Counterterms for Einstein-Gauss-Bonnet Gravity,''
  Class.\ Quant.\ Grav.\  {\bf 27}, 175014 (2010)
  [arXiv:0807.1256 [hep-th]].

  \bibitem{rhic}
J.~Adams {\it et al.}  [STAR Collaboration],
``Experimental and theoretical challenges in the search for the quark  gluon
plasma: The STAR collaboration's critical assessment of the  evidence from
RHIC collisions,''
Nucl.\ Phys.\  A {\bf 757}, 102 (2005) [arXiv:nucl-ex/0501009].

\bibitem{rhic2}
K.~Adcox {\it et al.}  [PHENIX Collaboration],
``Formation of dense partonic matter in relativistic nucleus nucleus
collisions at RHIC: Experimental evaluation by the PHENIX  collaboration,''
Nucl.\ Phys.\  A {\bf 757}, 184 (2005) [arXiv:nucl-ex/0410003].

\bibitem{lhc}
 Proceedings of Quark Matter 2011: 
J.\ Phys.\ GG\ {\bf 38}, number 12 (December 2011).

  \bibitem{fluid}
  E.~Shuryak,
  ``Why does the quark gluon plasma at RHIC behave as a nearly ideal fluid?,''
  Prog.\ Part.\ Nucl.\ Phys.\  {\bf 53}, 273 (2004) [arXiv:hep-ph/0312227]. 
 
\bibitem{fluid2}
 E.~V.~Shuryak,
  ``What RHIC experiments and theory tell us about properties of  quark-gluon
  plasma?,''
  Nucl.\ Phys.\  A {\bf 750}, 64 (2005) [arXiv:hep-ph/0405066].

\bibitem{reviewmateosetal} 
  J.~Casalderrey-Solana, H.~Liu, D.~Mateos, K.~Rajagopal and U.~A.~Wiedemann,
  ``Gauge/String Duality, Hot QCD and Heavy Ion Collisions,''
  arXiv:1101.0618 [hep-th].
  
  \bibitem{Rebhan} 
  A.~Rebhan and D.~Steineder,
  ``Violation of the Holographic Viscosity Bound in a Strongly Coupled Anisotropic Plasma,''
Phys.\ Rev.\ Lett.\  {\bf 108}, 021601 (2012)
[arXiv:1110.6825 [hep-th]].

\bibitem{mamo} 
  K.~A.~Mamo,
  ``Holographic Wilsonian RG Flow of the Shear Viscosity to Entropy Ratio in Strongly Coupled Anisotropic Plasma,''
  JHEP {\bf 1210}, 070 (2012)
  [arXiv:1205.1797 [hep-th]].
  
\bibitem{Chernicoff:2012iq} 
  M.~Chernicoff, D.~Fernandez, D.~Mateos and D.~Trancanelli,
  ``Drag force in a strongly coupled anisotropic plasma,''
  JHEP {\bf 1208}, 100 (2012)
  [arXiv:1202.3696 [hep-th]].
   
   \bibitem{giataganas} 
  D.~Giataganas,
  ``Probing strongly coupled anisotropic plasma,''
   JHEP {\bf 1207}, 031 (2012)
  [arXiv:1202.4436 [hep-th]].
  
  \bibitem{new_drag} 
  S.~Chakraborty and N.~Haque,
  ``Drag force in strongly coupled, anisotropic plasma at finite chemical potential,''
  arXiv:1410.7040 [hep-th].

 \bibitem{fadafan} 
  K.~B.~Fadafan and H.~Soltanpanahi,
  ``Energy loss in a strongly coupled anisotropic plasma,''
   JHEP {\bf 1210}, 085 (2012)
  [arXiv:1206.2271 [hep-th]].
  
 \bibitem{stopping} 
  B.~Muller and D.~-L.~Yan,
  ``Light Probes in a Strongly Coupled Anisotropic Plasma,''
 Phys.\ Rev.\ D {\bf 87}, 046004 (2013)
  [arXiv:1210.2095 [hep-th]].
  
\bibitem{Rebhan:2012bw} 
  A.~Rebhan and D.~Steineder,
  ``Probing Two Holographic Models of Strongly Coupled Anisotropic Plasma,''
   JHEP {\bf 1208}, 020 (2012)
  [arXiv:1205.4684 [hep-th]].
  
\bibitem{jet} 
  M.~Chernicoff, D.~Fernandez, D.~Mateos and D.~Trancanelli,
  ``Jet quenching in a strongly coupled anisotropic plasma,''
   JHEP {\bf 1208}, 041 (2012)
  [arXiv:1203.0561 [hep-th]].
  
  \bibitem{Chernicoff:2012bu} 
  M.~Chernicoff, D.~Fernandez, D.~Mateos and D.~Trancanelli,
  ``Quarkonium dissociation by anisotropy,''
  JHEP {\bf 1301}, 170 (2013)
  [arXiv:1208.2672 [hep-th]].
 
\bibitem{indians} 
  S.~Chakraborty and N.~Haque,
  ``Holographic quark-antiquark potential in hot, anisotropic Yang-Mills plasma,''
  Nucl.\ Phys.\ B {\bf 874}, 821 (2013)
  [arXiv:1212.2769 [hep-th]].
 
 \bibitem{Fadafan:2013bva} 
  K.~B.~Fadafan, D.~Giataganas and H.~Soltanpanahi,
  ``The Imaginary Part of the Static Potential in Strongly Coupled Anisotropic Plasma,''
 JHEP {\bf 1311}, 107 (2013)
  [arXiv:1306.2929 [hep-th]].

  \bibitem{langevin}
  D.~Giataganas and H.~Soltanpanahi,
  ``Universal properties of the Langevin diffusion coefficients,''
  Phys.\ Rev.\ D {\bf 89}, 026011 (2014)
  [arXiv:1310.6725 [hep-th]].
    
   \bibitem{langevin2}
  S.~Chakrabortty, S.~Chakraborty and N.~Haque,
  ``Brownian motion in strongly coupled, anisotropic Yang-Mills plasma: A holographic approach,''
 Phys.\ Rev.\ D {\bf 89}, 066013 (2014)
  [arXiv:1311.5023 [hep-th]].
  
  \bibitem{langevin3} 
  D.~Giataganas and H.~Soltanpanahi,
  ``Heavy Quark Diffusion in Strongly Coupled Anisotropic Plasmas,''
  JHEP {\bf 1406}, 047 (2014)
  [arXiv:1312.7474 [hep-th]].
  
\bibitem{Ali-Akbari:2013txa} 
  M.~Ali-Akbari and H.~Ebrahim,
  ``Chiral Symmetry Breaking: To Probe Anisotropy and Magnetic Field in QGP,''
 Phys.\ Rev.\ D {\bf 89}, no. 6, 065029 (2014)
  [arXiv:1309.4715 [hep-th]].
    
  \bibitem{photon} 
  L.~Patino and D.~Trancanelli,
  ``Thermal photon production in a strongly coupled anisotropic plasma,''
  JHEP {\bf 1302}, 154 (2013)
  [arXiv:1211.2199 [hep-th]].

  \bibitem{Wu:2013qja} 
  S.~-Y.~Wu and D.~-L.~Yang,
  ``Holographic Photon Production with Magnetic Field in Anisotropic Plasmas,''
 JHEP {\bf 1308}, 032 (2013)
  [arXiv:1305.5509 [hep-th]].
  
  \bibitem{Arciniega:2013dqa} 
  G.~Arciniega, P.~Ortega and L.~Patino,
  ``Brighter Branes, enhancement of photon production by strong magnetic fields in the gauge/gravity correspondence,''
  JHEP {\bf 1404}, 192 (2014)
  [arXiv:1307.1153 [hep-th]].
  
  \bibitem{dileptons}
   V.~Jahnke, A.~Luna, L.~Pati–o and D.~Trancanelli,
  ``More on thermal probes of a strongly coupled anisotropic plasma,''
  JHEP {\bf 1401}, 149 (2014)
  [arXiv:1311.5513 [hep-th]].
      
      \bibitem{chempot} 
  L.~Cheng, X.~H.~Ge and S.~J.~Sin,
  ``Anisotropic plasma with a chemical potential and scheme-independent instabilities,''
  Phys.\ Lett.\ B {\bf 734}, 116 (2014)
  [arXiv:1404.1994 [hep-th]].
      
      \bibitem{chempot1} 
  L.~Cheng, X.~H.~Ge and S.~J.~Sin,
  ``Anisotropic plasma at finite $U(1)$ chemical potential,''
  JHEP {\bf 1407}, 083 (2014)
  [arXiv:1404.5027 [hep-th]].

  \bibitem{Giataganas:2013lga} 
  D.~Giataganas,
  ``Observables in Strongly Coupled Anisotropic Theories,''
  PoS Corfu {\bf 2012}, 122 (2013)
  [arXiv:1306.1404 [hep-th]].
    
    \bibitem{seealso} 
  S.~Jain, N.~Kundu, K.~Sen, A.~Sinha and S.~P.~Trivedi,
  ``A Strongly Coupled Anisotropic Fluid From Dilaton Driven Holography,''
  arXiv:1406.4874 [hep-th].
    
      \bibitem{aneqc}
   M.~J.~Duff,
  ``Observations on Conformal Anomalies,''
  Nucl.\ Phys.\ B {\bf 125}, 334 (1977).
  
  \bibitem{aneqc1}
  S.~'i.~Nojiri and S.~D.~Odintsov,
  ``On the conformal anomaly from higher derivative gravity in AdS / CFT correspondence,''
  Int.\ J.\ Mod.\ Phys.\ A {\bf 15}, 413 (2000)
  [hep-th/9903033].
  
  \bibitem{aneqc2}
  M.~Blau, K.~S.~Narain and E.~Gava,
  ``On subleading contributions to the AdS / CFT trace anomaly,''
  JHEP {\bf 9909}, 018 (1999)
  [hep-th/9904179].

\bibitem{maldaedel} 
  X.~O.~Camanho, J.~D.~Edelstein, J.~Maldacena and A.~Zhiboedov,
  ``Causality Constraints on Corrections to the Graviton Three-Point Coupling,''
  arXiv:1407.5597 [hep-th].
  
  \bibitem{Brigante} 
  M.~Brigante, H.~Liu, R.~C.~Myers, S.~Shenker and S.~Yaida,
  ``Viscosity Bound Violation in Higher Derivative Gravity,''
  Phys.\ Rev.\ D {\bf 77}, 126006 (2008)
  [arXiv:0712.0805 [hep-th]].
  
  \bibitem{Mann} 
  M.~H.~Dehghani and R.~B.~Mann,
  ``Lovelock-Lifshitz Black Holes,''
  JHEP {\bf 1007}, 019 (2010)
  [arXiv:1004.4397 [hep-th]].
  
  \bibitem{ParkMann} 
  M.~Park and R.~B.~Mann,
  ``Deformations of Lifshitz holography with the Gauss-Bonnet term in ($n+1$) dimensions,''
  JHEP {\bf 1308}, 003 (2013)
  [arXiv:1305.5578 [hep-th]].
  
  \bibitem{reviewHR} 
  K.~Skenderis,
  ``Lecture notes on holographic renormalization,''
  Class.\ Quant.\ Grav.\  {\bf 19}, 5849 (2002)
  [hep-th/0209067].
  
  \bibitem{Hung:2011xb} 
  L.~Y.~Hung, R.~C.~Myers and M.~Smolkin,
  ``On Holographic Entanglement Entropy and Higher Curvature Gravity,''
  JHEP {\bf 1104}, 025 (2011)
  [arXiv:1101.5813 [hep-th]].
  
  \bibitem{emparan} 
  M.~M.~Caldarelli, R.~Emparan and B.~Van Pol,
  ``Higher-dimensional Rotating Charged Black Holes,''
  JHEP {\bf 1104}, 013 (2011)
  [arXiv:1012.4517 [hep-th]].
  
  \bibitem{other} 
  K.~B.~Fadafan,
  ``Charge effect and finite 't Hooft coupling correction on drag force and Jet Quenching Parameter,''
  Eur.\ Phys.\ J.\ C {\bf 68}, 505 (2010)
  [arXiv:0809.1336 [hep-th]].
  
\bibitem{other1} 
  V.~Jahnke, A.~S.~Misobuchi and D.~Trancanelli,
  ``The Chern-Simons diffusion rate from higher curvature gravity,''
  Phys.\ Rev.\ D {\bf 89}, 107901 (2014)
  [arXiv:1403.2681 [hep-th]].

\bibitem{KSS} 
  P.~Kovtun, D.~T.~Son and A.~O.~Starinets,
  ``Viscosity in strongly interacting quantum field theories from black hole physics,''
  Phys.\ Rev.\ Lett.\  {\bf 94}, 111601 (2005)
  [hep-th/0405231].
  
    \bibitem{beyond} 
  Y.~Kats and P.~Petrov,
  ``Effect of curvature squared corrections in AdS on the viscosity of the dual gauge theory,''
  JHEP {\bf 0901}, 044 (2009)
  [arXiv:0712.0743 [hep-th]].
  
  \bibitem{beyond0} 
  M.~Brigante, H.~Liu, R.~C.~Myers, S.~Shenker and S.~Yaida,
  ``The Viscosity Bound and Causality Violation,''
  Phys.\ Rev.\ Lett.\  {\bf 100}, 191601 (2008)
  [arXiv:0802.3318 [hep-th]].
  
  \bibitem{beyond05} 
  X.~H.~Ge, Y.~Matsuo, F.~W.~Shu, S.~J.~Sin and T.~Tsukioka,
  ``Viscosity Bound, Causality Violation and Instability with Stringy Correction and Charge,''
  JHEP {\bf 0810}, 009 (2008)
  [arXiv:0808.2354 [hep-th]].
  
  \bibitem{beyond1}
  A.~Buchel, R.~C.~Myers and A.~Sinha,
  ``Beyond eta/s = 1/4 pi,''
  JHEP {\bf 0903}, 084 (2009)
  [arXiv:0812.2521 [hep-th]].
  
  \bibitem{beyond2}
  R.~-G.~Cai, Z.~-Y.~Nie, N.~Ohta and Y.~-W.~Sun,
``Shear Viscosity from Gauss-Bonnet Gravity with a Dilaton Coupling,''
Phys.\ Rev.\ D {\bf 79} (2009) 066004
[arXiv:0901.1421 [hep-th]].

\bibitem{beyond23} 
  X.~H.~Ge and S.~J.~Sin,
  ``Shear viscosity, instability and the upper bound of the Gauss-Bonnet coupling constant,''
  JHEP {\bf 0905}, 051 (2009)
  [arXiv:0903.2527 [hep-th]].

\bibitem{beyond25} 
  X.~H.~Ge, S.~J.~Sin, S.~F.~Wu and G.~H.~Yang,
  ``Shear viscosity and instability from third order Lovelock gravity,''
  Phys.\ Rev.\ D {\bf 80}, 104019 (2009)
  [arXiv:0905.2675 [hep-th]].

  \bibitem{beyond3}
    J.~de Boer, M.~Kulaxizi and A.~Parnachev,
  ``AdS(7)/CFT(6), Gauss-Bonnet Gravity, and Viscosity Bound,''
  JHEP {\bf 1003}, 087 (2010)
  [arXiv:0910.5347 [hep-th]].
  
  \bibitem{beyond4}
  X.~O.~Camanho, J.~D.~Edelstein and M.~F.~Paulos,
  ``Lovelock theories, holography and the fate of the viscosity bound,''
  JHEP {\bf 1105}, 127 (2011)
  [arXiv:1010.1682 [hep-th]].
  
  \bibitem{beyond5}
  T.~Takahashi and J.~Soda,
  ``Pathologies in Lovelock AdS Black Branes and AdS/CFT,''
  Class.\ Quant.\ Grav.\ {\bf 29}, 035008 (2012)
  [arXiv:1108.5041 [hep-th]].
  
  \bibitem{sera} 
  S.~Cremonini,
  ``The Shear Viscosity to Entropy Ratio: A Status Report,''
  Mod.\ Phys.\ Lett.\ B {\bf 25}, 1867 (2011)
  [arXiv:1108.0677 [hep-th]].
    
\bibitem{Iqbal} 
  N.~Iqbal and H.~Liu,
  ``Universality of the hydrodynamic limit in AdS/CFT and the membrane paradigm,''
  Phys.\ Rev.\ D {\bf 79}, 025023 (2009)
  [arXiv:0809.3808 [hep-th]].
  
  \bibitem{anis_superfluid} 
  A.~Bhattacharyya and D.~Roychowdhury,
  ``Viscosity bound for anisotropic superfluids in higher derivative gravity,''
  arXiv:1410.3222 [hep-th].
  
  \bibitem{riccati1} 
  I.~Papadimitriou and K.~Skenderis,
  ``Correlation functions in holographic RG flows,''
  JHEP {\bf 0410}, 075 (2004)
  [hep-th/0407071].
  
  \bibitem{riccati2} 
  I.~Papadimitriou and A.~Taliotis,
  ``Riccati equations for holographic 2-point functions,''
  JHEP {\bf 1404}, 194 (2014)
  [arXiv:1312.7876 [hep-th]].
  
  \bibitem{chinese_new}
L.~Cheng, X.-H.~Ge, Z.-Y.~Sun,
``Thermoelectric DC conductivities with momentum dissipation from higher derivative gravity,'' 
arXiv:1411.5452 [hep-th].  
   
\bibitem{SS} 
  D.~T.~Son and A.~O.~Starinets,
  ``Minkowski space correlators in AdS / CFT correspondence: Recipe and applications,''
  JHEP {\bf 0209}, 042 (2002)
  [hep-th/0205051].
   
  \bibitem{PSS}
  G.~Policastro, D.~T.~Son and A.~O.~Starinets,
  ``From AdS/CFT correspondence to hydrodynamics,''
  JHEP {\bf 0209}, 043 (2002)
  [arXiv:hep-th/0205052].
    
  \bibitem{PSS2} 
  G.~Policastro, D.~T.~Son and A.~O.~Starinets,
  ``From AdS/CFT correspondence to hydrodynamics. 2. Sound waves,''
  JHEP {\bf 0212}, 054 (2002)
  [hep-th/0210220].
  
  \bibitem{KS} 
  P.~K.~Kovtun and A.~O.~Starinets,
  ``Quasinormal modes and holography,''
  Phys.\ Rev.\ D {\bf 72}, 086009 (2005)
  [hep-th/0506184].

\bibitem{Kiritsis} 
  U.~G\"ursoy, I.~Iatrakis, E.~Kiritsis, F.~Nitti and A.~O'Bannon,
  ``The Chern-Simons Diffusion Rate in Improved Holographic QCD,''
  JHEP {\bf 1302}, 119 (2013)
  [arXiv:1212.3894 [hep-th]].
    
\end{thebibliography}
\end{document}